\documentclass[11pt]{article}
\pdfoutput=1
\usepackage{jheppub}
\usepackage{amsmath}
\usepackage{bm}
\usepackage{slashed}
\usepackage{graphicx}
\usepackage{tabularx}
\usepackage{diagbox}
\usepackage{slashed}
\usepackage{float}

\newlength{\dummysp}
\newcommand{\beq}{\begin{eqnarray}}
\newcommand{\eeq}{\end{eqnarray}}

\newcommand{\gappeq}{\mathrel{\rlap {\raise.5ex\hbox{$>$}}
{\lower.5ex\hbox{$\sim$}}}}
\newcommand{\lappeq}{\mathrel{\rlap{\raise.5ex\hbox{$<$}}
{\lower.5ex\hbox{$\sim$}}}}

\newcommand{\ben}{\begin{enumerate}}
\newcommand{\een}{\end{enumerate}}

\newcommand{\bit}{\begin{itemize}}
\newcommand{\eit}{\end{itemize}}

\def\[{\left [}
\def\]{\right ]}
\def\({\left (}
\def\){\right )}

\title{Condensates and anomaly cascade in vector-like theories}
   
\author{Mohamed M. Anber}
\affiliation{Department of Physics, Lewis $\&$ Clark College, Portland, OR 97219, USA}
\emailAdd{manber@lclark.edu}

\abstract{We study the bilinear and higher-order fermion condensates in $4$-dimensional $SU(N)$ gauge theories with a single Dirac fermion in a general representation. Augmented with a mixed anomaly between the $0$-form discrete chiral, $1$-form center, and $0$-form baryon number symmetries (BC anomaly), we sort out theories that admit  higher-order condensates and vanishing fermion bilinears. Then, the BC anomaly is utilized to prove, in the absence of a topological quantum field theory, that nonvanishing fermion bilinears are inevitable in infrared-gapped theories with $2$-index (anti)symmetric fermions. We also contrast the BC anomaly with the $0$-form anomalies and show that it is the former anomaly that determines the infrared physics;  we argue that the BC anomaly  lurks deep to the infrared while the $0$-form anomalies are just variations of local terms. We provide evidence of this assertion by studying the BC anomaly in vector-like theories compactified on a small spacial circle.  These theories are weakly-coupled, under analytical control, and they admit a dual description in terms of abelian photons that determine the deep infrared dynamics. We show that the dual photons talk directly to the $1$-form center symmetry in order to match the BC anomaly, while the $0$-form anomalies are variations of local terms and are matched by fiat. Finally, we  study the fate of the BC anomaly in the compactified theories when they are held at a finite temperature. The effective field theory that describes the low-energy physics is $2$-dimensional. We show that the BC anomaly cascades from $4$ to $2$ dimensions.}

\begin{document}

\maketitle

\flushbottom

\section{Introduction}

't Hooft anomaly matching conditions is one of the very few handles on the nonperturbative phenomena in strongly-coupled theories \cite{tHooft:1979rat}. The anomaly is an unremovable phase in the partition function that needs to be matched between the ultraviolet (UV) and infrared (IR), which imposes constraints on the viable scenarios of the phases  of a given asymptotically-free gauge theory that flows to strong coupling in the IR.   Recently, it has been realized that the class of 't Hooft anomalies is larger than what has been known since the 80s. It was discovered in \cite{Gaiotto:2014kfa,Gaiotto:2017yup} that Higher-form symmetries may also become anomalous, which can be used to impose further constraints on strongly-coupled theories. These original papers were followed by a plethora of other works that attempted to use the new anomalies to study various aspects of quantum field theory, see \cite{Komargodski:2017smk,Sulejmanpasic:2018upi,Wan:2018bns,Cordova:2019jnf,Cordova:2019uob,Anber:2019nfu,Cordova:2019jqi,Anber:2020qzb,Cordova:2018acb,Anber:2018iof,Anber:2018jdf,Anber:2018xek,Unsal:2020yeh,Sulejmanpasic:2020zfs,Tanizaki:2019rbk,Cherman:2020cvw,Cherman:2019hbq,Shimizu:2017asf,Brennan:2020ehu}  for a non-comprehensive list. 

One can understand the new development as an anomaly of a global transformation on the field content in the background of a fractional topological charge, an 't Hooft flux \cite{tHooft:1979rtg,vanBaal:1982ag}, of the center symmetry of the gauge group.  This anomaly was further enlarged in \cite{Anber:2019nze} by considering the most general fractional charges in the baryon number, color,  and flavor (BCF) directions. This anomaly was dubbed the BCF anomaly (or only BC anomaly when we have a single flavor), and was also studied in \cite{Anber:2020gig} on nonspin manifolds. One of the profound consequences of the BCF anomaly is the deconfinement of quarks on axion domain walls, a phenomenon that is attributed to an intertwining between the light (axion) and heavy (hadron) degrees of freedom at  the core of the domain wall. The intertwining between the different degrees of freedom can also have an important effect on models of axion inflation \cite{Anber:2020qzb}. 

In this paper we consider a $4$-dimensional asymptotically-free $SU(N)$ gauge theory with a single Dirac flavor $\Psi$ in a general representation ${\cal R}$ and strong-coupling scale $\Lambda$. The theory admits a $U(1)_B$ baryon and  $\mathbb Z^{d\chi}_{2T_{\cal R}}$ discrete chiral symmetries, where $T_{\cal R}$ is the Dynkin index of the representation. As the theory flows to the IR and enters its strongly-coupled regime, we assume that it forms a nonvanishing bilinear fermion condensate $\langle \bar\Psi \Psi \rangle \neq 0$. Then,  the discrete chiral symmetry breaks spontaneously, $\mathbb Z^{d\chi}_{2T_{\cal R}}\rightarrow \mathbb Z_2$, leaving behind $T_{\cal R}$ degenerate vacua. These vacua are separated by domain walls of width $\sim\Lambda^{-1}$. If the bilinear fermion condensate vanishes, then higher-order condensates may form, which, in general, break  $\mathbb Z^{d\chi}_{2T_{\cal R}}$ down to a discrete subgroup.  We ponder on several questions:
\begin{enumerate}
\item A theory with an 't Hooft anomaly precludes a unique gapped vacuum. What do anomalies inform us about the breaking of $\mathbb Z^{d\chi}_{2T_{\cal R}}$? Is there an anomaly that grants the full breaking of  $\mathbb Z^{d\chi}_{2T_{\cal R}}$ down to $\mathbb Z_2$? Is this anomaly unique or there are several anomalies that yield the same result? Is one of the anomalies more restricting than the others, and does this depend on ${\cal R}$?

\item How do the domain walls respond to these anomalies?

\item  How are the anomalies matched at finite temperature?

\end{enumerate}

 Indeed, it is well-known that a vector-like theory admits a mixed anomaly between $\mathbb Z^{d\chi}_{2T_{\cal R}}$ and $U(1)_B$, we denote it by $\mathbb Z^{d\chi}_{2T_{\cal R}}\left[U(1)_B\right]^2$, which needs to be matched between the UV and IR. If the bilinear condensate forms, then the existence of $T_{\cal R}$ degenerate vacua will automatically match the anomaly.  Sometimes, however, a  $T_{\cal R}$ degeneracy is an overkill in the sense that only a subset of $T_{\cal R}$ vacua are needed for the matching. This happens if the anomaly $\mathbb Z^{d\chi}_{2T_{\cal R}}\left[U(1)_B\right]^2$ gives a phase valued in a proper subgroup of  $\mathbb Z^{d\chi}_{T_{\cal R}}$. In this case we might set $\langle \bar\Psi \Psi \rangle =0$ and argue that higher-order condensates  break the chiral symmetry to a subgroup that  gives the exact number of vacua needed to match the anomaly.  For example, $SU(4)$ with a Dirac fermion in the $2$-index symmetric representation has $T_{\cal R}=6$ and we expect that the bilinear condensate, if it forms,  breaks $\mathbb Z^{d\chi}_{12}$ spontaneously resulting in $6$ vacua. The $\mathbb Z^{d\chi}_{2T_{\cal R}}\left[U(1)_B\right]^2$ anomaly, however, is valued in $\mathbb Z_3$ and can be matched by $3$, instead of, $6$ vacua. Then, it is a plausible scenario, in the light of the $\mathbb Z^{d\chi}_{2T_{\cal R}}\left[U(1)_B\right]^2$ anomaly, that the bilinear condensate vanishes and the four-fermion condensate $\langle\bar \Psi\Psi \bar \Psi\Psi\rangle$ forms and yields $3$ vacua.    
 
 Another anomaly that gives the exact same conclusion is $\mathbb Z^{d\chi}_{2T_{\cal R}}\left[\mbox{gravity}\right]^2$, which results from the action of $\mathbb Z^{d\chi}_{2T_{\cal R}}$ on the fermions in the gravitational background of a nonspin manifold.

Given this classical result, one wonders whether a yet-to-be-discovered anomaly may impose a stronger constraint on the number of the degenerate vacua and gives us a nonperturbative exact statement about this number. We address this question in the light of the BC anomaly and show that it provides constraints stronger than or equal to the constraints from the traditional  $\mathbb Z^{d\chi}_{2T_{\cal R}}\left[U(1)_B\right]^2$  and $\mathbb Z^{d\chi}_{2T_{\cal R}}\left[\mbox{gravity}\right]^2$  anomalies. In particular, we show, in the absence of a topological quantum field theory, that $SU(4k)$ with fermions in the $2$-index (anti)symmetric representation has to break its discrete chiral symmetry down to the fermion number $\mathbb Z_2$ and yields exactly $T_{4k\pm2}$ vacua. Thus, the BC anomaly excludes the above mentioned  four-fermi condensate scenario. 

In fact, we examined all $SU(N)$, with $3\leq N\leq 9$, asymptotically-free gauge theories with fermions in a general representation and concluded that there are only two types of theories that have a stronger response to the BC anomaly than the traditional anomalies. These theories are: (i)  $SU(4k)$ with fermions in the $2$-index symmetric representation and (ii)  $SU(4k)$, $k>1$, with fermions in the $2$-index antisymmetric representation . Nonetheless, we shall argue that it is the BC anomaly, in fact, that ``orders" the breaking of the discrete chiral symmetry.   We show that a  domain wall that separates two distinct vacua couples to a $3$-form field $a^{(3)}$ that transforms non-trivially under a $2$-form symmetry, which is at the heart of the BC anomaly. $a^{(3)}$, however, is inert under both $ \mathbb Z^{d\chi}_{2T_{\cal R}}$ and $U(1)_B$. This observation seems to suggest that $\mathbb Z^{d\chi}_{2T_{\cal R}}\left[U(1)_B\right]^2$ anomaly is matched by ``fiat". In the rest of the paper we provide a  justification of this hypothesis. 

Because of the strong-coupling nature of the $4$-dimensional theory, it is extremely hard to provide a detailed analysis of what really happens in its vacuum; there is no separation of scales and all phenomena, e.g., confinement and chiral symmetry breaking, take places at the same scale $\sim \Lambda$.  In order to test our hypothesis, we study the fate of anomalies in a semi-classical setup. We compactify the vector-like theories on a small circle $\mathbb S^1_L$ of circumference $L$, such that $\Lambda L\ll 1$, and give the fermions  periodic boundary conditions on $\mathbb S^1_L$. This is not a thermal theory; the periodic boundary conditions turn the thermal partition function into a graded-state sum. We say that the theory lives on $\mathbb R^3\times \mathbb S^1_L$. In addition,  we add adjoint massive fermions or a double-trace deformation in order to force the theory into its weakly-coupled semi-classical regime, without spoiling the original global symmetry. Effectively, the IR theory lives in $3$ dimensions, it abelianizes, and becomes amenable to analytical studies. We can also go to a dual (magnetic) description, where the ``dual photons" play the main role in determining the pattern of the discrete chiral symmetry breaking.  We show that the dual photons couple nontrivially to the higher-form symmetry, and therefore, the BC anomaly is communicated from the UV to the deep IR. The $\mathbb Z^{d\chi}_{2T_{\cal R}}\left[U(1)_B\right]^2$ anomaly, on the other hand, shows up as a variation of a  local action and does not talk to the photons. In this sense, we say that $\mathbb Z^{d\chi}_{2T_{\cal R}}\left[U(1)_B\right]^2$ anomaly is matched by fiat. This analysis provides evidence that it is the BC anomaly that talks to the IR degrees of freedom. Our work uses and generalizes the observation that was first made by Poppitz and  Wandler \cite{Poppitz:2020tto} that cubic- and mixed-$U(1)$ anomalies are matched by local background-field-dependent topological terms instead of chiral-Lagrangian Wess-Zumino-Witten terms, while the $1$-form center symmetry talks directly to the dual photons.   We further study in detail the $SU(4k)$ theory on $\mathbb R^3\times \mathbb S^1_L$ with $2$-index (anti)symmetric fermions and analyze the dynamics that leads to the full breaking of $\mathbb Z_{2(4k\pm 2)}^{d\chi}$, the expected result in accordance with the BC anomaly. As a byproduct, we identify new composite instantons that play a major role in the IR. 

We also examine the fate of the BC anomaly as we heat up the the theory. The strong coupling nature of the $4$-dimensional theory hinders our ability to answer this question. We circumvent this difficulty, again, by studying the compactified theory at a finite temperature. Now, in addition to the spacial circle $\mathbb S^1_L$, we also have a thermal circle $\mathbb S^1_\beta$, where $\beta$ is the inverse temperature, and we say that the theory lives on $\mathbb R^2\times \mathbb S^1_L\times \mathbb S^1_\beta$. Effectively, it can be shown that the theory is dual to a $2$-dimensional electric-magnetic Coulomb gas. We do not attempt to solve the effective $2$-dimensional theory since the strong-coupling problem might resurrect near the confinement/deconfinement transition. However, we trace the fate of the BC anomaly on  $\mathbb R^2\times \mathbb S^1_L\times \mathbb S^1_\beta$ and show that this anomaly ``cascades" from $4$ down to $2$-dimensions.  We also use renormalization group equations to argue that the theory admits flat directions in the dual photon space as we heat it up, and eventually the long-range force of the dual photons, which were responsible in the first place for the breaking of the chiral symmetry, is tamed  indicating that the chiral symmetry is restored. In this case we find that the BC anomaly becomes ``confined", or in other words local, and is matched by fiat.     

This paper is organized as follows. In Section \ref{vector like theories} we review the symmetries and the corresponding background fields in $4$-dimensional vector-like theories with a single Dirac fermion in a general representation. We also review the essence of the BC anomaly and compare it to the traditional anomalies. Next, we  study the condensates and the role of the BC anomaly. In Section \ref{Vector-like theories on a circle} we work out  the construction of the vector-like theories on a small circle; we consider both the perturbative and nonperturbative aspects and we introduce the dual theory. Then, we show in great details how the BC anomaly is reproduced in the dual picture and argue that it lurks deep in the IR.  This is in contradistinction with the traditional anomalies, since they are realized as the variation of   local actions  that do not communicate with the IR degrees of freedom. We also trace the fate of the BC anomaly at we heat up the dual theory. In Section \ref{Examples on the circle} we work out the details of $SU(4k)$ on the small circle with fermions in the $2$-index (anti)symmetric representation and identify the microscopic objects that are responsible for the full breaking of the discrete chiral symmetry. Finally, we consider these theories at a finite temperature and use renormalization group equations to understand the realization of the BC anomaly as we heat up the system.

\section{Vector-like theories on $\mathbb R^4$}
\label{vector like theories}

\subsection{Symmetries and background fields}

We consider $SU(N)$ Yang-Mills theory endowed with a single left-handed massless Weyl fermion $\psi$ in a representation  ${\cal R}$ along with another left-handed massless Weyl fermion $\tilde \psi$ transforming in the complex conjugate representation. Collectively, we can also talk about a single Dirac fermion in $\cal R$. The $4$-dimensional Lagrangian reads
\begin{eqnarray}
{\cal L}_4=-\frac{1}{4g^2} \mbox{tr}_F\left[F_{MN}F^{MN}\right]+\frac{\theta}{32\pi^2}\mbox{tr}_F \left[F_{MN} \tilde F^{MN}\right]+i \bar \psi \bar \sigma^M D_M \psi+i \bar {\tilde\psi} \bar \sigma^M D_M \tilde\psi\,,
\end{eqnarray}
where $M,N=0,1,2,3$ and the partition function $\cal Z$ is defined over a large closed manifold.  
The Dynkin index of the representation is denoted by $T_{\cal R}$ (we use the normalization $\mbox{tr}_F[T^aT^b]=\delta_{ab}$, where $T^a$ are the generators of the Lie-algebra) and its dimension is $\mbox{dim}_{{\cal R}}$. Strictly speaking, since the fermions are massless, we could rotate the the $\theta$ angle away by applying a chiral transformation on $\psi$ and $\tilde \psi$. Keeping the topological term, however, will serve a later purpose.    The theory admits  the  global symmetry:
\begin{eqnarray}
G^{\scriptsize\mbox{Global}}=\mathbb Z_{2T_{\cal R}}^{d\chi}\times \frac{U(1)_B}{\mathbb Z_{N/p}\times \mathbb Z_2}\times \mathbb Z^{(1)}_{p}\,,
\end{eqnarray}
where $\mathbb Z_2$ is the fermion number (which is a subgroup of the Lorentz group, and hence, we mod it out), $p=\mbox{gcd}(N,n)$, and $n$ is the N-ality\footnote{The N-ality of a representation is the number of boxes in the Young tabulate mod $N$.} of  ${\cal R}$. Notice that $\mathbb Z_{N/p}$, which is a subgroup of the center group $\mathbb Z_N$,  acts faithfully on the fermions, and therefore,  we needed to mod it out since it is part of the gauge group.  $\mathbb Z_{2T_{\cal R}}^{d\chi}$ and $U(1)_B$ are respectively the $0$-form discrete chiral and baryon number symmetries acting on $\psi$ and $\tilde \psi$:
\begin{eqnarray}
\mathbb Z_{2T_{\cal R}}^{d\chi}: \psi\rightarrow e^{i\frac{2\pi}{2T_{\cal R}}}\psi, \tilde\psi\rightarrow e^{i\frac{2\pi}{2T_{\cal R}}}\tilde\psi\,, \quad U(1)_B:  \psi\rightarrow e^{i\alpha}\psi, \tilde\psi\rightarrow e^{-i\alpha}\tilde\psi\,.
\end{eqnarray}
 Finally, $\mathbb Z^{(1)}_{p}$, provided that $p>1$, is the $1$-form symmetry that acts on the fundamental Wilson's loops.
 
 When the representation is real, then we slightly modify the above procedure since in this case it is enough to have a single fermion without the need to introduce another fermion transforming in the would-be complex-conjugate representation. We use the symbol $\lambda$ for the real Weyl fermions.  For example, a single adjoint Weyl fermion defines super Yang-Mills theory with $T_{\scriptsize\mbox{adj}}=2N$, $\mbox{dim}_{\scriptsize\mbox{adj}}=N^2-1$, and  global symmetry $\mathbb Z^{d\chi}_{2N}\times \mathbb Z_N^{(1)}$. 
 
We also need to turn on background fields of  $G^{\scriptsize\mbox{Global}}$ since they play a pivotal role in determining 't Hooft anomalies. Introducing a background field of $U(1)_B$ is straight forward; we just include it in the covariant derivative. Thus, we write $D=d+i A -iV^{(1)}$, where $A$ is the color gauge field and its field strength is $F=dA+A\wedge A$, and $V^{(1)}$ is the $1$-form $U(1)_B$ gauge field with  field strength   $F^{B(2)}=dV^{(1)}$. Introducing background fields of discrete symmetries is more involved. In order to turn on a background field of  the discrete  chiral symmetry $\mathbb Z^{d\chi}_{2T_{\cal R}}$, we introduce a pair of $0$-form and $1$-form fields $\left(b^{(0)}, B^{(1)}\right)$ that satisfy the relation $2T_{\cal R}B^{(1)}=d b^{(0)}$ and demand that the integral of the $1$-form field $db^{(0)}$ over $1$-cycles is in $\mathbb Z$, i.e., $\oint db^{(0)}=2\pi \mathbb Z$, which in turn implies $\oint B^{(1)}\in \frac{2\pi}{2T_{\cal R}}\mathbb Z$, where the integral of $B^{(1)}$ is performed over $1$-cycles. These fields are also invariant under the gauge transformation $B^{(1)}\rightarrow B^{(1)}+d\omega^{(0)}$ and $b^{(0)}\rightarrow b^{(0)}+2T_{\cal R}\omega^{(0)}$, and $d\omega^{(0)}$ has quantized periods over $1$-cycles: $\oint d \omega^{(0)}\in 2\pi\mathbb Z$. One may think of $b^{(0)}$ as the phase of a charge-$2T_{\cal R}$ non-dynamical Higgs field that acquires a vacuum expectation value and breaks a $U(1)$  gauge field down to the $\mathbb Z_{2T_{\cal R}}$ discrete  field $B^{(1)}$.  Under the transformation $\psi\rightarrow e^{i\frac{b^{(0)}}{2T_{\cal R}}}\psi$ and  $\tilde \psi\rightarrow e^{i\frac{b^{(0)}}{2T_{\cal R}}}\tilde \psi$ the measure acquires a phase $e^{i\int\frac{b^{(0)}}{32\pi^2}\mbox{tr}_F \left[F_{MN} \tilde F^{MN}\right]}$. Therefore, following the analysis of \cite{Poppitz:2020tto}, one can think of $b^{(0)}$ as a background $\theta$ angle, and we shall use the former instead of the latter in the following discussion. 
 
Next, we turn to the $\mathbb Z_N$ center group of $SU(N)$. As we mentioned above, only a $\mathbb Z_{N/p}$, $p=\mbox{gcd}(N,n)$, subgroup of the center acts faithfully on the fermions, leaving behind a global $\mathbb Z_p$ that we may choose to turn on a background field associated to it. Yet, one can excite a background field of the full center $\mathbb Z_N$ owning to the baryon symmetry. The simplest way to understand this assertion is by examining the transition functions ${\cal G}_{ij}$ on the overlap between two patches $U_i$ and $U_j$ that cover the $4$-dimensional manifold. On the overlap $U_i \cap U_j$ we have
\begin{eqnarray}
\psi_i={\cal G}_{ij} \psi_j\,, \quad {\cal G}_{ij}={\cal G}_{ij}^{\mathbb Z_N}{\cal G}^{U(1)_B}_{ij}\,,
\end{eqnarray}
where ${\cal G}_{ij}^{\mathbb Z_N}$ and ${\cal G}^{U(1)_B}_{ij}$ are respectively the transition functions of the center and baryon number symmetries. A similar transformation holds for $\tilde \psi$. The consistency of the gauge theory requires that the transition functions satisfy the following cocycle condition 
\begin{eqnarray}
{\cal G}_{ij}{\cal G}_{jk}{\cal G}_{ki}=1
\label{cocycle condition}
\end{eqnarray}
on the triplet overlap $U_i \cap U_j\cap U_k$. The most general solution of the cocycle condition is obtained by taking ${\cal G}^{\mathbb Z_N}_{ij}=e^{i 2\pi\frac{n}{N}}$ and ${\cal G}^{U(1)_B}_{ij}=e^{-i 2\pi\frac{n}{N}}$, where the additional factor of $n$ that appears in the exponent in ${\cal G}^{\mathbb Z_N}_{ij}$ accounts for the fact that the fermions transform in a representation of N-ality $n$. This explains why one can always excite the full $\mathbb Z_N$ background. Indeed, when $p>1$, then one may not use $U(1)_B$ and  instead choose to turn on a background field of $\mathbb Z_p\subset \mathbb Z_N$. As it turns out, exciting the full $\mathbb Z_{N}$ will impose stronger constrains on the theory by employing the related 't Hooft anomalies. 

The background field of $\mathbb Z_N$ is an 't Hooft flux that carries a fractional topological charge.  The modern way of thinking of 't Hooft fluxes is via higher-form symmetries, as was done in \cite{Anber:2020xfk}. From now on, we consider $\mathbb Z_N^{(1)}$ $1$-form symmetry, which in principle acts on Wilson's loops. In order to turn on a background  field of $\mathbb Z_N^{(1)}$ we use a pair of $1$-form and $2$-form fields $\left(B^{c(2)}, B^{c(1)}\right)$ such that $NB^{c(2)}=dB^{c(1)}$, see \cite{Kapustin:2014gua}. The periods of $B^{c(1)}$ are quantized in multiples of $2\pi$: $\oint dB^{c(1)}\in 2\pi \mathbb Z$, where the integral is over $2$-cycles. Now, owing to the relation $NB^{c(2)}=dB^{c(1)}$, we obtain $\oint B^{c(2)}\in \frac{2\pi }{N}\mathbb Z$. Next,  we define the  $U(N)$ connection $\tilde A\equiv A+\frac{B^{c(1)}}{N}I_{N\times N}$ with gauge field strength $\tilde F=d\tilde A+\tilde A\wedge \tilde A$. The field strength $\tilde F$ satisfies the relation $\mbox{tr}_F \tilde F=dB^{c(1)}=N B^{c(2)}$.  Going from $SU(N)$ to $U(N)$ introduces a non-physical extra degree of freedom. In  order to eliminate this degree of freedom, we postulate the following invariance $\tilde A \rightarrow \tilde A +\lambda^{(1)}$ under the $1$-form gauge field  $\lambda^{(1)}$. Subsequently,  the pair  $\left(B^{c(2)}, B^{c(1)}\right)$ transforms as $B^{c(2)}\rightarrow B^{c(2)} +d\lambda^{(1)}$ and $B^{c(1)}\rightarrow B^{c(1)}+N\lambda^{(1)}$, such that the relation  $NB^{c(2)}=dB^{c(1)}$ remains intact. The covariant derivative of the matter field is obtained by replacing $A$ with $\tilde A$, i.e.,  $D=d+i\tilde A -iV^{(1)}$. The invariance of $D$ under $\lambda^{(1)}$ enforces  the baryon background field to transform as $V\rightarrow V+n\lambda^{(1)}$, where the factor of $n$ is the N-ality of the representation (recall the discussion after the cocycle condition (\ref{cocycle condition})), and hence, we find that $F^B$ transforms as $F^B \rightarrow F^B+nd\lambda^{(1)}$. 

\subsection{The baryon-color (BC) 't Hooft anomaly}

Turning on the baryon and the center background fields enables us to find the most general perturbative 't Hooft anomaly on a spin manifold. As was shown in \cite{Anber:2020xfk}, this is an 't Hooft anomaly of the discrete chiral symmetry in the background of both $\mathbb Z_N^{(1)}$ and $U(1)_B$ fields, and hence, the name   baryon-color (BC) 't Hooft anomaly. Succinctly, we can compute the anomaly from the triangle diagrams with vertices sourced by the following $2$-form combinations  $\tilde F-B^{c(2)}$ and $F^B-nB^{c(2)}$, which are invariant under the $1$-form gauge transformation with parameter $\lambda^{(1)}$.  The triangle diagrams yield the following color and baryon number topological densities:
\begin{eqnarray}
\nonumber
q^c=\frac{1}{8\pi^2}\left[\mbox{tr}_F\left(\tilde F \wedge \tilde F\right)-N B^{c(2)}\wedge B^{c(2)} \right]\,, \quad q^B=\frac{1}{8\pi^2}\left[F^B-n B^{c(2)}\right]  \wedge \left[F^B-n B^{c(2)}\right]\,. \\
\label{topological charge densities}
\end{eqnarray}
Then, we perform a discrete chiral transformation in the background of the BC background to find that the partition function ${\cal Z}$ acquires the phase:
\begin{eqnarray}
{\cal Z}\xrightarrow{\mathbb Z^{d\chi}_{2T_{\cal R}}} e^{i\frac{2\pi}{T_{\cal R}}\left(T_{\cal R}Q^c+\mbox{dim}_{\cal R}Q^B\right)}  {\cal Z}\,,
\label{BC anomaly on R4}
\end{eqnarray}
where $Q^c=\int q^c$ and $Q^B=\int q^B$ and the integral is performed over a closed $4$-dimensional spin manifold. Owing to the facts: $\frac{1}{8\pi^2}\int \tilde F \wedge \tilde F\in \mathbb Z$,  $\frac{1}{8\pi^2}\int  F^B \wedge F^B\in \mathbb Z$, and  $\frac{N}{8\pi^2}\int B^{c(2)}\wedge B^{c(2)}\in \frac{1}{N}\mathbb Z$, we find $Q^c=1-\frac{1}{N}$ and $Q^B=(\ell+\frac{n}{N})^2$, $\ell \in \mathbb Z$.  Since $T_{\cal R}Q^c+\mbox{dim}_{\cal R}Q^B$ is the Dirac-index, which is always an integer, then the phase of the partition function in the BC background is valued in $\mathbb Z_{T_{\cal R}}$ or a subgroup of it:
\begin{eqnarray}
\mbox{ BC Anomaly}= e^{i\frac{2\pi}{T_{\cal R}}\left(T_{\cal R}Q^c+\mbox{dim}_{\cal R}Q^B\right)} \in \mathbb Z_{T_{\cal R}}\,.
\end{eqnarray}
 At this stage one might think that the BC anomaly does not impose on the dynamics any further constraints beyond the traditional anomalies\footnote{The most refined phase of the $\mathbb Z_{2T_{\cal R}}^{d\chi}\left[\mbox{gravity}\right]^2$ anomaly comes from a calculation on a nonspin manifold. Fermions are ill-defined when the manifold is nonspin, e.g. $\mathbb{CP}^2$. In order to render the fermions well-defined on $\mathbb{CP}^2$, we turn on a monopole background of $U(1)_B$ with charge $\frac{1}{2}$. The fractional  monopole flux combines with the fractional flux of the gravitational  $\mathbb{CP}^2$ instanton and yields an integer Dirac index $=1$. Hence, one immediately finds the anomaly in (\ref{traditional anomaly}).} $\mathbb Z_{2T_{\cal R}}^{d\chi}\left[U(1)_B\right]^2$ and $\mathbb Z_{2T_{\cal R}}^{d\chi}\left[\mbox{gravity}\right]^2$, since the latter are also valued in $\mathbb Z_{T_{\cal R}}$:
 \begin{eqnarray}
 \mathbb Z_{2T_{\cal R}}^{d\chi}\left[U(1)_B\right]^2=\mathbb Z_{2T_{\cal R}}^{d\chi}\left[\mbox{gravity}\right]^2=e^{i 2\pi  \frac{\mbox{dim}_{\cal R}}{T_{\cal R}}} \in \mathbb Z_{T_{\cal R}} \,.
 \label{traditional anomaly}
 \end{eqnarray}
  However, as we will argue in the next section, unlike the $\mathbb Z_{2T_{\cal R}}^{d\chi}\left[U(1)_B\right]^2$ and  $\mathbb Z_{2T_{\cal R}}^{d\chi}\left[\mbox{gravity}\right]^2$ anomalies,  the BC anomaly  is more restrictive and communicates non-trivial information to the low-energy confining phase deep in the IR. This will be evident in the semi-classical analysis  that we will perform on the theory upon compactifying it  on a small circle. It is also worth mentioning that one may compute the BC anomaly in a nonspin background, as was done in \cite{Anber:2020gig}. We checked, however, that the BC anomaly on a nonspin manifold does not impose more restrictions on the condensates compared to the same anomaly on a spin manifold. 

Finally, let us note that when $p=\mbox{gcd}(N,n)>1$, then we can also turn on the background of $\mathbb Z_p^{(1)}\subset \mathbb Z_N^{(1)}$ without the need to employ $U(1)_B$. This can be accomplished by constraining the quantization of $B^{c(2)}$ over $2$-cycles to obey $\oint B^{c(2)}=\frac{2\pi}{p}\mathbb Z$, and hence,  $Q^c=\frac{N}{8\pi^2}\int B^{c(2)}\wedge B^{c(2)}\in \frac{N}{p^2}\mathbb Z$ .   Then, we encounter a mixed 't Hooft anomaly between $\mathbb Z_{2T_{\cal R}}^{d\chi}$ and $\mathbb Z_{p}^{(1)}$, which gives the phase
\begin{eqnarray}
{\cal Z}\xrightarrow{\mathbb Z^{d\chi}_{2T_{\cal R}}} e^{i\frac{2\pi}{T_{\cal R}}\left(T_{\cal R} Q^c\right)}  {\cal Z}=e^{i2\pi  \frac{N}{p^2}}  {\cal Z}\,,
\label{gauging center symmetry}
\end{eqnarray}
which is less restrictive than the phase from the BC anomaly.

\subsection{Condensates and role of the BC anomaly}
\label{Condensates and role of the BC anomaly}

As we flow to the IR, the theory may or may not break its discrete chiral symmetry. In the following, we assume that: (1)  the theory generates a mass gap and the discrete chiral symmetry breaks, which can be probed via the non-vanishing color-singlet bilinear condensate $\langle\psi\tilde \psi \rangle$ or higher-order condensates, (2) the is no topological quantum field theory accompanying the IR phase\footnote{The possibility of IR topological quantum field theory was considered in \cite{Cordova:2019jqi,Cordova:2019bsd}.}, and (3) the theory does not form massless composite fermions in the IR. The formation of the condensates, then, implies that in general the full or partial  breaking of  $\mathbb Z^{d\chi}_{2T_{\cal R}}$ takes place, leading to $T_{\cal R}$ or fewer distinct vacua. The conclusion about the full breaking of $\mathbb Z^{d\chi}_{2T_{\cal R}}$ cannot be guaranteed unless there is an anomaly that is valued in $\mathbb Z_{T_{\cal R}}$ and not only in a proper subgroup of it. Only in this case the saturation of the anomaly in the IR, indeed, demands the full breaking of $\mathbb Z_{2T_{\cal R}}^{d\chi}$.  

If $\mbox{gcd}(T_{\cal R}, \mbox{dim}_{\cal R})>1$, then (\ref{traditional anomaly}) implies that  both $\mathbb Z_{2T_{\cal R}}^{d\chi}\left[U(1)_B\right]^2$ and $\mathbb Z_{2T_{\cal R}}^{d\chi}\left[\mbox{gravity}\right]^2$ anomalies do not necessarily demand the full breaking of the chiral symmetry; the partial breaking $\mathbb Z^{d\chi}_{2T_{\cal R}}\rightarrow \mathbb Z_{2\mbox{gcd}(T_{\cal R}, \mbox{dim}_{\cal R})}$ is sufficient to match the anomalies. Similarly, when $\mbox{gcd}(T_{\cal R}, \left(T_{\cal R}Q^c+\mbox{dim}_{\cal R}Q^B\right))>1$, then the BC anomaly can be matched via the breaking $\mathbb Z^{d\chi}_{2T_{\cal R}}\rightarrow \mathbb Z_{2\mbox{gcd}\left(T_{\cal R}, \left(T_{\cal R}Q^c+\mbox{dim}_{\cal R}Q^B\right)\right)}$.

\begin {table}
\begin{center}
\tabcolsep=0.11cm
\footnotesize
\begin{tabular}{|c|c|c|c|c|c|c|}
\hline
Group &$\cal R$ & $T_{\cal R}$ & $\mbox{dim}_{\cal R}$ & $\mathbb Z_{2T_{\cal R}}^{d\chi}\left[U(1)_B\right]^2$ & BC & Condensate\\\hline
$SU(3)$& $(2,0)$ & $5$ & $6$ & $\mathbb  Z_5$ & $\mathbb Z_5$ & $\left\langle \tilde\psi\psi \right\rangle$ \\\hline
&$(3,0)$ & $15$ & $10$ & $\mathbb  Z_3$ & $\mathbb Z_3$ & $\left\langle (\tilde\psi\psi)^5 \right\rangle$\\\hline
&$(1,1)$ & $6$ & $8$ & --& $\mathbb Z_3$ & $\left\langle \lambda\lambda \right\rangle$\\\hline
&$(2,1)$ & $20$ & $15$ & $\mathbb  Z_4$ & $\mathbb Z_4$ & $\left\langle (\tilde\psi\psi)^5 \right\rangle$\\\hline
\hline\hline
$SU(4)$&$(2,0,0)$ & $6$ & $10$ & $\mathbb  Z_3$ & $\mathbb Z_6$ & $\left\langle \tilde\psi\psi \right\rangle$ \\\hline
&$(3,0,0)$ & $21$ & $20$ & $\mathbb  Z_{21}$ & $\mathbb Z_{21}$ & $\left\langle \tilde\psi\psi \right\rangle$ \\\hline
&$(0,1,0)$ & $2$ & $6$ & -- & $1$ & No constraints \\\hline
&$(0,2,0)$ & $16$ & $20$ & -- & $\mathbb Z_4$ & $\left\langle (\lambda\lambda)^2  \right\rangle$\\\hline
&$(1,0,1)$ & $8$ & $15$ & -- & $\mathbb Z_{4}$ & $\left\langle \lambda\lambda  \right\rangle$ \\\hline
&$(1,1,0)$ & $13$ & $20$ & $\mathbb  Z_{13}$ & $\mathbb Z_{13}$ & $\left\langle \tilde\psi\psi \right\rangle$ \\\hline
&$(2,0,1)$ & $33$ & $36$ & $\mathbb  Z_{11}$ & $\mathbb Z_{11}$ & $\left\langle (\tilde\psi\psi)^3 \right\rangle$\\\hline
\hline\hline
$SU(5)$ & $(2,0,0,0)$ & $7$ & $15$ & $\mathbb Z_7$ & $\mathbb Z_7$ & $\left\langle \tilde\psi\psi \right\rangle$\\\hline
& $(0,1,0,0)$ & $3$ & $10$ & $\mathbb Z_3$ & $\mathbb Z_3$ & $\left\langle \tilde\psi\psi \right\rangle$\\\hline
& $(1,0,0,1)$ & $10$ & $24$ & -- & $\mathbb Z_5$ & $\left\langle \lambda\lambda  \right\rangle$\\\hline
& $(1,1,0,0)$ & $22$ & $40$  & $\mathbb Z_{11}$ & $\mathbb Z_{11}$ & $\left\langle (\tilde\psi\psi)^2 \right\rangle$\\\hline
& $(1,0,1,0)$ & $24$ & $45$ & $\mathbb Z_8$ & $\mathbb Z_{8}$ & $\left\langle (\tilde\psi\psi)^3 \right\rangle$\\\hline
\end{tabular}
\caption{ \label{breaking pattern} The asymptotically free representations of $SU(3)$ to $SU(5)$. We use the Dynkin labels to designate the representation: ${\cal R}=(n_1,n_2,...,n_{N-1})\equiv \sum_{a=1}^{N-1}n_a \bm w_a$, where $\bm w_a$ are the fundamental weights. A representation is said to be real if $(n_1,n_2,...,n_{N-1})=(n_{N-1},n_{N-2},..., n_1)$.  For example,  $(1,1)$, $(1,0,1)$, $(0,1,0)$, $(0,2,0)$ are all real representations. In this case, one needs to be more careful  since $U(1)_B$ is enhanced to  $SU(2)_f$ flavor symmetry. We avoid this extra complication by considering a single Weyl fermion, $\lambda$, whenever the representation is real. Then,  the discrete chiral symmetry becomes $\mathbb Z_{T_{\cal R}}^{d\chi}$ and the baryon number symmetry  as well as the anomaly $\mathbb Z_{2T_{\cal R}}^{d\chi}\left[U(1)_B\right]^2$ disappear. Notice that we exclude the defining representation $(1,0,0,...,0)$ since theories with fundamentals do not have genuine discrete chiral symmetries. In the next to last column we list the phases of both $\mathbb Z_{2T_{\cal R}}^{d\chi}\left[U(1)_B\right]^2$ (which is equal to $\mathbb Z_{2T_{\cal R}}^{d\chi}\left[\mbox{gravity}\right]^2$ anomaly) and BC anomalies. In the last column we display the higher-order condensate that saturates the BC anomaly.}
\end{center}
\end{table}

\begin {table}
\begin{center}
\tabcolsep=0.11cm
\footnotesize
\begin{tabular}{|c|c|c|c|c|c|c|}
\hline
Group &$\cal R$ & $T_{\cal R}$ & $\mbox{dim}_{\cal R}$ & $\mathbb Z_{2T_{\cal R}}^{d\chi}\left[U(1)_B\right]^2$ & BC & Condensate\\\hline
$SU(6)$ & $(2,0,0,0,0)$ & $8$ & $21$ & $\mathbb Z_8$ & $\mathbb Z_8$ & $\left\langle \tilde\psi\psi \right\rangle$\\\hline
& $(0,1,0,0,0)$ & $4$ & $15$ & $\mathbb Z_4$ & $\mathbb Z_4$ & $\left\langle \tilde\psi\psi \right\rangle$\\\hline
& $(0,0,1,0,0)$ & $6$ & $20$ & $\mathbb Z_3$ & $\mathbb Z_3$ & $\left\langle (\lambda\lambda)^2  \right\rangle$\\\hline
& $(1,0,0,0,1)$ & $12$ & $35$ &-- & $\mathbb Z_{6}$ & $\left\langle \lambda\lambda  \right\rangle$  \\\hline
& $(1,1,0,0,0)$ & $33$ & $70$ & $\mathbb Z_{33}$ & $\mathbb Z_{33}$ & $\left\langle \tilde\psi\psi \right\rangle$\\\hline
\hline\hline
$SU(7)$ & $(2,0,0,0,0,0)$ & $9$ & $28$ & $\mathbb Z_9$ & $\mathbb Z_9$ & $\left\langle \tilde\psi\psi \right\rangle$\\\hline
& $(0,1,0,0,0,0)$ & $5$ & $21$ & $\mathbb Z_5$ & $\mathbb Z_5$ & $\left\langle \tilde\psi\psi \right\rangle$\\\hline
 & $(1,0,0,0,0,1)$ & $14$ & $48$ & -- & $\mathbb Z_7$ & $\left\langle \lambda\lambda  \right\rangle$\\\hline
& $(0,0,1,0,0,0)$ & $10$ & $35$ & $\mathbb Z_2$ & $\mathbb Z_2$ & $\left\langle (\tilde\psi\psi)^5 \right\rangle$\\\hline
\hline\hline
$SU(8)$ & $(2,0,0,0,0,0,0)$ & $10$ & $36$ & $\mathbb Z_5$ & $\mathbb Z_{10}$  & $\left\langle \tilde\psi\psi \right\rangle$\\\hline
& $(0,1,0,0,0,0,0)$ & $6$ & $28$ & $\mathbb Z_3$ & $\mathbb Z_6$  & $\left\langle \tilde\psi\psi \right\rangle$ \\\hline
& $(0,0,0,1,0,0,0)$ & $20$ & $70$ & $\mathbb Z_2$ & $\mathbb Z_2$ & $\left\langle (\lambda\lambda)^5  \right\rangle$\\\hline
& $(1,0,0,0,0,0,1)$ & $16$ & $63$ & --& $\mathbb Z_{8}$ &$\left\langle \lambda\lambda  \right\rangle$ \\\hline
& $(0,0,1,0,0,0,0)$ & $15$ & $56$ & $\mathbb Z_{15}$ & $\mathbb Z_{15}$  & $\left\langle \tilde\psi\psi \right\rangle$\\\hline
\hline\hline
$SU(9)$ & $(2,0,0,0,0,0,0,0)$ & $11$ & $45$ & $\mathbb Z_{11}$ & $\mathbb Z_{11}$ & $\left\langle \tilde\psi\psi \right\rangle$\\\hline
& $(0,1,0,0,0,0,0,0)$ & $7$ & $36$ & $\mathbb Z_{7}$ & $\mathbb Z_7$ & $\left\langle \tilde\psi\psi \right\rangle$\\\hline
& $(1,0,0,0,0,0,0,1)$ & $18$ & $80$ & -- & $\mathbb Z_9$ & $\left\langle \lambda\lambda  \right\rangle$\\\hline
\end{tabular}
\caption{ \label{breaking pattern 2} The asymptotically free representations of $SU(6)$ to $SU(8)$. For details see the caption of Table \ref{breaking pattern}. Notice that the first non-vanishing condensate in the representation $(0,0,1,0,0)$ of $SU(6)$ is a $4$-fermion operator since the fermion bilinear vanishes identically for group theory reasons, see \cite{Yamaguchi:2018xse,Anber:2019nfu}. }
\end{center}
\end{table}

In Tables \ref{breaking pattern} and \ref{breaking pattern 2}  we display the asymptotically free representations of $SU(N)$, $3\leq N\leq 9$, gauge theories as well as their anomalies. When the representation is real, then we limit the analysis to a single Weyl fermion and in this case the discrete chiral symmetry is $\mathbb Z^{d\chi}_{T_{\cal R}}$ instead of $\mathbb Z^{d\chi}_{2T_{\cal R}}$. Also, in this case the BC anomaly is reduced to the phase given by (\ref{gauging center symmetry}).

For all complex representations, except two cases,  we find that  both $\mathbb Z_{2T_{\cal R}}^{d\chi}\left[U(1)_B\right]^2$ and BC anomalies yield the same phase. The exceptions are: 
\begin{itemize}
\item $SU(4k)$ theories with fermions in the $2$-index symmetric representation: ${\cal R}=(2,0,...,0)$ with $T_{\cal R}=4k+2$ and $\mbox{dim}_{\cal R}=2k(4k+1)$.
\item $SU(4k)$, $k>1$, theories  with fermions in the  $2$-index anti-symmetric representation: ${\cal R}=(0,1,0,...,0)$  with $T_{\cal R}=4k-2$ and $\mbox{dim}_{\cal R}=2k(4k-1)$.
\end{itemize}
Here we find $\mbox{gcd}(T_{\cal R}, \mbox{dim}_{\cal R})=2$, while $\left(T_{\cal R}Q^c+\mbox{dim}_{\cal R}Q^B\right)_{(2,0,...,0)}=4k+3$,\\ $\left(T_{\cal R}Q^c+\mbox{dim}_{\cal R}Q^B\right)_{(0,1,...,0)}=4k-1$, and hence,  $\mbox{gcd}(T_{\cal R}, \left(T_{\cal R}Q^c+\mbox{dim}_{\cal R}Q^B\right))=1$, making the BC anomaly more restricting than $\mathbb Z_{2T_{\cal R}}^{d\chi}\left[U(1)_B\right]^2$ and $\mathbb Z_{2T_{\cal R}}^{d\chi}\left[\mbox{gravity}\right]^2$ anomalies. Then, the BC anomaly demands the full breaking of $\mathbb Z_{2(4k\pm 2)}^{d\chi}$ and the formation of $4k\pm2$ distinct vacua, in the symmetric and antisymmetric representations, respectively. Notice that both of these representations admit a $\mathbb Z^{(1)}_2$ symmetry acting on Wilson's loop and gauging it  leads to a trivial phase, as can be easily seen from  (\ref{gauging center symmetry}). We conclude, in the absence of a topological quantum field theory, that nonvanishing fermion bilinears are inevitable in infrared-gapped $SU(N)$ gauge theories with $2$-index (anti)symmetric fermions.

We also observe that when the phase of the BC anomaly is in a prober subgroup of the discrete chiral symmetry, then a plausible scenario is that  the bilinear condensate vanishes and higher-order condensates form. In the last column  of Tables \ref{breaking pattern} and \ref{breaking pattern 2} we display the possible higher-order  condensate that saturates the BC anomaly.  For example, the discrete chiral symmetry of $SU(4)$ Yang-Mills theory with a single Dirac fermion in the $(2,0,1)$ representation is $\mathbb Z_{66}^{d\chi}$ and the formation of the bilinear condensate suggests that the theory admits $33$ vacua in the IR. However, the  BC anomaly can be matched via the breaking $\mathbb Z_{66}^{d\chi}\rightarrow \mathbb Z_{6}$, suggesting that an IR phase with only $11$ vacua is enough to match the anomaly. Thus, a plausible scenario that matches the anomalies is the vanishing of both the bilinear and four-fermion condensates $\langle \tilde \psi \psi \rangle=\langle \tilde \psi \psi \tilde \psi \psi \rangle=0$ and the formation of the six-fermion condensate $\left\langle (\tilde\psi\psi)^3 \right\rangle\equiv\langle \tilde \psi \psi \tilde \psi \psi  \tilde \psi \psi  \rangle\neq0$. 

The exceptional cases discussed above give us an insight into the special role of the BC anomaly compared to the traditional anomalies $\mathbb Z_{2T_{\cal R}}^{d\chi}\left[U(1)_B\right]^2$ and $\mathbb Z_{2T_{\cal R}}^{d\chi}\left[\mbox{gravity}\right]^2$.  We argue that  it is the BC anomaly that lurks deep in the IR and demands the existence of multiple vacua. In Section (\ref{Vector-like theories on a circle}) we put this hypothesis into test by studying the same theory on a small circle. This setup enables us to perform semi-classical calculations and examine various phenomena that are rather difficult, if not impossible, to understand in the strong-coupling regime. In particular, we will show that it is the BC anomaly that influence the IR dynamics, while the $\mathbb Z_{2T_{\cal R}}^{d\chi}\left[U(1)_B\right]^2$ anomaly  is the variation of a local action and is matched by fiat, but otherwise does not influence the IR dynamics.

Before delving into the analysis on the circle, let us show how the BC anomaly is matched in $4$-dimensions deep in the IR.  As the condensate forms, domain walls will interpolate between the degenerate vacua. Let $a^{(3)}$ be the $3$-form field that couples to the domain wall such that $\oint a^{(3)}\in 2\pi \mathbb Z$ and the integral is over $3$-cycles. Then, one can write down the following $5$-dimensional Wess-Zumino-Witten term that matches the anomaly in the IR:
\begin{eqnarray}
\nonumber
S_{WZW}=\int_W d\omega^{(0)}\wedge\left[da^{(3)}-\frac{N}{8\pi^2}B^{c(2)}\wedge B^{c(2)}+\frac{\mbox{dim}_{\cal R}}{8\pi^2 T_{\cal R}}\left[F^B-n B^{c(2)}\right]  \wedge \left[F^B-n B^{c(2)}\right] \right]\,.\\
\label{mod WZW}
\end{eqnarray}
Under a $\mathbb Z_{2T_{\cal R}}^{d\chi}$ transformation we use $\oint d\omega^{(0)}\in 2\pi \mathbb Z$ and find $e^{-i\delta S_{WZW}}\in \mathbb Z_{T_{\cal R}}$. A closer examination of the action (\ref{mod WZW}) reveals some important information about the IR physics that cannot be seen without the BC anomaly. As we discussed above, the $2$-form field transforms under the $1$-form gauge field $\lambda^{(1)}$ as: $B^{c(2)}\rightarrow B^{c(2)}+d\lambda^{(1)}$. This, in turn,  demands that the $3$-form field  transforms as
\begin{eqnarray}
a^{(3)}\rightarrow a^{(3)}+\frac{N}{2\pi}B^{c(2)}\wedge \lambda^{c(1)}+\frac{N}{4\pi}\lambda^{c(1)}\wedge d\lambda^{(1)}\,,
\end{eqnarray}
which indicates that the $\mathbb Z^{(1)}_N$ $1$-form symmetry lurks deep in the IR and affects the domain wall dynamics. Let us contrast this behavior with the traditional $\mathbb Z_{2T_{\cal R}}^{d\chi}\left[U(1)_B\right]^2$ anomaly. Here, all we need to do is to turn off  $B^{c(2)}$ in (\ref{mod WZW}). Then, we still find that $e^{-i\delta S_{WZW}}\in \mathbb Z_{T_{\cal R}}$. However, the the $3$-form field that couples to the domain wall does not transform under $U(1)_B$ or $ \mathbb Z^{d\chi}_{2T_{\cal R}}$; the $\mathbb Z_{2T_{\cal R}}^{d\chi}\left[U(1)_B\right]^2$ anomaly is matched trivially. 

Although our analysis in $4$ dimensions might sound like an academic exercise due to the lack of any control on the strong dynamics, in the next section we show how  our reasoning becomes manifest once we push the theory into its weakly-coupled regime.

\section{Vector-like theories on $\mathbb R^3\times \mathbb S^1_L$}
\label{Vector-like theories on a circle}

\subsection{Perturbative aspects}

In this section we study the vector-like theories by compactifying the $x^3$ direction on a small circle $\mathbb S^1_L$ with circumference $L$ and demand that $\Lambda$, the strong coupling scale of the theory,  is taken such that $L\Lambda\ll 1$. In addition, the fermions obey periodic boundary conditions on $\mathbb S^1_L$.  This setup guarantees that the theory enters its semi-classical regime, and hence, we can use reliable analytical methods to analyze it. We say that the theory lives on $\mathbb R^3\times \mathbb S^1_L$.  Further, the analysis of the theory simplifies considerably if we force it into its center-symmetric point (more on that will be discussed below). This can be achieved either by adding a double-trace deformation
\begin{eqnarray}
{\cal L}_{\scriptsize \mbox{DT}}=\sum_j c_j \left|\mbox{tr}_F\left(e^{ij\oint A_3}\right)\right|^2\,,
\end{eqnarray}
with large positive coefficients $c_i$,
 or by adding massive adjoint  fermions with mass $\sim L^{-1}$. Both of these two alterations to the theory neither change its global symmetries nor its 't Hooft anomalies. However, we note that, depending on ${\cal R}$, adding adjoint fermions might not achieve the goal of stabilizing the theory at the center symmetric point, as was discussed in details in \cite{Anber:2017pak}.

This construction was considered before in a plethora of works, and we refer the reader to the literature for more details, see \cite{Dunne:2016nmc} for a review. Here, it suffices to say that the theory completely abelianizes at the center-symmetric point:  $SU(N)$ breaks  down spontaneously  to the maximal abelian subgroup $U(1)^{N-1}$. Now, all fields that appear in the low-energy effective Lagrangian are valued in the Cartan subalgebra space, which we label by bold face symbols. At energy scales much smaller than the inverse circle radius we dimensionally reduce the theory to $3$ dimensions with effective Lagrangian:
\begin{eqnarray}
{\cal L}_3=-\frac{L}{4g^2}\bm F_{\mu\nu} \bm F^{\mu\nu}-\frac{b^{(0)}}{8\pi^2}\epsilon^{\alpha\mu\nu}\partial_\alpha\bm\Phi\cdot\bm F_{\mu\nu}+\frac{1}{2g^2 L}\partial_\mu\bm \Phi \partial^\mu \bm \Phi+V(\bm \Phi)+{\cal L}_{3,f}\,,
\label{low energy lag}
\end{eqnarray}
where $\mu,\nu=0,1,2$ and in our notation, for example, $\bm \Phi=(\phi_1,\phi_2,..., \phi_{N-1})$. The field $\bm\Phi$ is the gauge field holonomy in the $\mathbb S^1_L$ direction: $L \bm A_3\equiv \bm \Phi$. The second term is the $4$-dimensional topological term dimensionally reduced to $3$ dimensions. As we promised above, we traded the $\theta$ angle for the background field $b^{(0)}$ of the discrete chiral symmetry.  The potential $V(\bm \Phi)$ is the Gross-Pisarski-Yaffe (GPY) potential \cite{Gross:1980br}, which results from summing towers of  the Kaluza-Klein excitations of the gauge field, the ${\cal R}$ fermions, and the massive adjoint  fermions. We always force $V(\bm \Phi)$ to be minimized at the center-symmetric point either by  adding massive adjoint fermions  or double-trace  deformation. The center-symmetric value of $\bm \Phi$ is
\begin{eqnarray}
\bm \Phi_0=\frac{2\pi \bm \rho}{N}\,,
\end{eqnarray}
where $\bm \rho=\sum_{a=1}^{N-1} \bm w_a$ is the Weyl vector and $\bm w_a$ are the fundamental weights. See the discussion immediately before (\ref{color background decomp}) for more comments on the meaning of the center-symmetric vacuum.   The holonomy fluctuations about $\bm \Phi_0$ have masses of order $\sim \frac{g}{L}$, and thus, we can neglect them whenever we are interested in energies much smaller than $\frac{g}{L}$. The $U(1)^{N-1}$ gauge fields $\bm F_{\mu\nu}$ are given, as usual, by $\bm F_{\mu\nu}=\partial_\mu \bm A_\nu -\partial_\nu \bm A_\mu$. Both $d\bm \Phi$ and $\bm F$ satisfy the quantization conditions $\oint d\bm \Phi \in2\pi \bm \alpha_a\mathbb Z$ and  $\oint \bm F\in2\pi \bm \alpha_a\mathbb Z$, where the integrals are  taken respectively over $1$- and $2$-cycles,  for all simple roots $\bm\alpha_a$, $a=1,2,...,N-1$. The simple roots have length $\bm \alpha^2_a=2$ and  satisfy the relation $\bm \alpha_a\cdot\bm w_b=\delta_{ab}$.

Finally, we comment on the fermion term in (\ref{low energy lag}). The $3$-dimensional fermion Lagrangian is given by (here we consider the Lagrangian of $\psi$. An identical Lagrangian holds for $\tilde \psi$)
\begin{eqnarray}
{\cal L}_{3,f}=i\sum_{\bm \mu \in {\cal R}}\sum_{p\in \mathbb Z}\bar \psi_p^{\bm\mu} \left[\bar \sigma^\mu\left(\partial_\mu+i \bm A_\mu\cdot \bm \mu\right)+i\bar \sigma^3\left(\frac{2\pi p}{L} +\frac{\bm \mu \cdot \bm\Phi}{L}\right)\right]\psi_p^{\bm\mu}\,,
\label{fermion lag}
\end{eqnarray}
where $\bm\mu$ are the weights of ${\cal R}$ and $p$ is the Kaluza-Klein index. The effective $3$-dimensional fermion mass can be readily found from (\ref{fermion lag}): $M_{p\,,\bm \mu}=|\frac{2\pi p}{L} +\frac{\bm \mu \cdot \bm\Phi}{L}|$. This mass has to be non-vanishing for every non-zero value of $\bm\mu$, otherwise the low-energy $U(1)^{N-1}$ gauge theory becomes strongly coupled, which in turn, invalidates any semi-classical treatment. Yet, in certain situations, depending on ${\cal R}$, nonperturbative effects (these are monopole instantons and/or their composites) can give the fermions a small $4$-dimensional Dirac mass, rendering the theory IR safe. Alternatively, we can also turn on a holonomy of $U(1)_B$, which ensures that all the fermions are massive with mass $\sim L^{-1}$. To this end, we decompose the $4$-dimensional $U(1)_B$ $1$-form background field as
\begin{eqnarray}
V^{(1)}=V_{3D}^{(1)}+ \left(\frac{\kappa}{L}+V_{\mathbb S_L^1}^{(0)}\right)\frac{dx^3}{L}\,,
\label{baryon decomp}
\end{eqnarray}
where  $V_{3D}^{(1)}$ is the $1$-form background field in $\mathbb R^3$, $\frac{\kappa}{L}$ is the $U(1)_B$ holonomy in the $\mathbb S^1_L$ direction,  and $V^{(0)}_{\mathbb S_L^1}$ are the holonomy fluctuations. Turning on $\kappa$ modifies the $3$-dimensional  fermion mass to $M_{p\,,\bm \mu}=|\frac{2\pi p}{L} +\frac{\bm \mu \cdot \bm\Phi-\kappa}{L}|$, and now all the fermions are massive with mass $\sim L^{-1}$.

\subsection{The dual Lagrangian}

We shall investigate the realization of the symmetries as well as the  BC anomaly on $\mathbb R^3 \times \mathbb S^1_L$ in the semi-classical regime deep in the IR. In order achieve this goal we need to utilize a dual (magnetic) description.   To this end, we introduce the  dual photon $\bm \sigma$ as a Lagrange multiplier that enforces the Bianchi identity $\epsilon^{\alpha\mu\nu}\partial_\alpha\bm F_{\mu\nu}=0$. We augment the Lagrangian (\ref{low energy lag}) with the term $\frac{1}{4\pi} \epsilon^{\alpha \mu\nu}\bm \sigma \cdot \partial_\alpha \bm F_{\mu\nu}$ and then vary the combination with respect to $\bm F_{\mu\nu}$ to find:
\begin{eqnarray}
\bm F_{\mu\nu}=-\frac{g^2}{2\pi L}\epsilon_{\mu\nu \alpha}\left[\partial^\alpha \bm \sigma+ \frac{b^{(0)}}{2\pi}\partial^\alpha \bm \Phi \right]\,.
\label{F in terms of sigma}
\end{eqnarray}
Next, we break $\bm \Phi$ into two parts: the vacuum  $\bm\Phi_0$ and the fluctuations around it $\bm\varphi$ such that $\bm \Phi=\bm \Phi_0+\bm \varphi$ . Substituting (\ref{F in terms of sigma}) into (\ref{low energy lag}) we then find 
\begin{eqnarray}
\nonumber
{\cal L}_3=\frac{g^2}{8\pi^2 L}\left(\partial_\alpha \bm \sigma+\frac{ b^{(0)}}{2\pi}\partial_\alpha \bm \varphi \right)\cdot \left(\partial^\alpha \bm \sigma+  \frac{b^{(0)}}{2\pi}\partial^\alpha \bm \varphi \right)+\frac{1}{2g^2L}\partial_\alpha \bm \varphi\cdot \partial^\alpha \bm \varphi+V(\bm \Phi)+{\cal L}_{3,f}\,.\\
\label{sigma Lag}
\end{eqnarray}
The domain of $\bm\sigma$ can be determined as follows. The integral of $d\bm \sigma$ over $1$-cycles is equal to the electric charge enclosed by the cycles. Since all the electric  charges are allowed probe charges when the group is $SU(N)$, then the domain of $\bm\sigma$ is the finest lattice, which is the weight lattice: $\oint d\bm \sigma\in 2\pi \bm w_a\mathbb Z$ for all $a=1,2,...,N-1$.

Under a discrete chiral transformation $b^{(0)}$ transforms as $b^{(0)}\rightarrow b^{(0)}+2\pi$, Then, the invariance of (\ref{sigma Lag}) under $\mathbb Z^{d\chi}_{2T_{\cal R}}$ demands that the dual photons shift as
\begin{eqnarray}
\bm \sigma \rightarrow \bm\sigma -\bm \varphi-\bm C\,,
\label{sigma transformation}
\end{eqnarray}
where $\bm C$ is a constant vector that belongs to the weight lattice, which is allowed  owing to the fact that it is the derivatives of $\bm \sigma$ and $\bm \varphi$ that appear in (\ref{sigma Lag}). The constant $\bm C$ can be unambiguously determined once we take the the nonperturbative effects into account. 

\subsection{Nonperturbative aspects}

The theory also admits monopole-instantons. The action of the lowest Kaluza-Klein monopoles ($p=0$ monopoles, where $p$ is the Kaluza-Klein index) is 
\begin{eqnarray}
S_{\bm \alpha_a}=\frac{4\pi}{g^2}\bm \alpha_a\cdot \bm \Phi_0
\label{monopole action}
\end{eqnarray}
for every simple root $\bm \alpha_a$, $a=1,2,...,N-1$. There is also one extra monopole instanton that corresponds to the affine root $\bm\alpha_N=-\sum_{a=1}^{N-1} \bm \alpha_a$ with an action $S_{\bm \alpha_N}=\frac{8\pi^2}{g^2}+\frac{4\pi}{g^2}\bm \alpha_N\cdot \bm \Phi_0$. Module ${\cal O}(1)$ normalization coefficients, the 't Hooft vertex associated with each monopole, including the affine monopole $a=N$, is given by:
\begin{eqnarray}
{\cal M}_{a}=e^{\left(-\frac{8\pi^2}{g^2}+ib^{(0)}\right)\delta_{Na}}e^{-\frac{4\pi}{g^2}\bm \alpha_a\cdot \bm \Phi_0}e^{i \bm \alpha_a\cdot \left(\bm \sigma+\frac{b^{(0)}}{2\pi}\bm\varphi\right)}(\psi \tilde \psi)^{I_a}\,, \quad a=1,2,...,N\,.
\label{Monopole vortices}
\end{eqnarray}
 The exponent $I_a$ is the Callias index that counts the number of the fermion zero modes in the background of the monopole \cite{Callias:1977kg,Poppitz:2008hr}. The index  of the lowest Kaluza-Klein monopole is given by \cite{Anber:2014lba,Anber:2017pak}:
\begin{eqnarray}
I_a=\sum_{\bm \mu}\left \lfloor \frac{\bm \Phi_0\cdot \bm \mu}{2\pi} \right\rfloor \bm \alpha_a\cdot \bm \mu\,, a=1,2,...,N-1\,, \quad\quad  I_N=2T_{\cal R}-\sum_{a=1}^{N-1} I_{a}\,.
\label{index theorem}
\end{eqnarray}

Each monopole vertex has to respect the global symmetries. First, it is evident that  ${\cal M}_{a}$ is invariant under $U(1)_B$. Next, in order to respect the invariance under $\mathbb Z^{d\chi}_{2T_{\cal R}}$ we express the constant $\bm C$ in (\ref{sigma transformation}) as a general vector in the weight lattice as 
\begin{eqnarray}
\bm C=2\pi \sum_{a=1}^{N-1}{\cal K}_a \bm w_a\,.
\label{C}
\end{eqnarray}
 Then, the invariance of each vertex under $\mathbb Z^{d\chi}_{2T_{\cal R}}$ fixes the values of ${\cal K}_a$:
\begin{eqnarray}
{\cal K}_a=\frac{I_a}{T_{\cal R}}\,.
\end{eqnarray}

As we shall discuss below, in some cases the lowest Kaluza-Klein monopoles are insufficient to construct the full low-energy effective potential $V(\bm\sigma)$. Thus, we need to turn into the first excited monopoles. Their actions can be obtained from (\ref{monopole action}) by replacing $\bm \Phi_0\rightarrow \bm \Phi_0 +\pi\bm \alpha_a$ \cite{Davies:2000nw}: 
\begin{eqnarray}
S^{p=1}_{\bm \alpha_a}=\frac{8\pi^2}{g^2}+\frac{4\pi}{g^2}\bm \alpha_a\cdot \bm \Phi_0\,.
\label{composite monopole}
\end{eqnarray}
 This action suggests that a $p=1$ monopole can be thought of as a composite configuration of the original monopole plus a Belavin-Polyakov-Schwarz-Tyupkin  (BPST) instanton\footnote{The action of a BPST instanton is $\frac{8\pi^2}{g^2}$ and can be thought of as the composite of all the monopoles that are charged under the simple and affine roots. Therefore, a BPST instanton has a total number of $2T_{\cal R}$  fermion zero modes.}. The number of the fermion zero modes in the background of the excited monopoles can be read from (\ref{index theorem}) after adding   $T_{\cal R}$ extra zero modes of $\psi$ and $T_{\cal R}$ extra zero modes of $\tilde\psi$.

The proliferation of monopoles or monopole-composites will lead to confinement and chiral symmetry breaking. Several examples that illustrate the important points of this paper will be worked out in later sections.

\subsection{The BC anomaly on $\mathbb R^3\times \mathbb S^1_L$}

Next, we turn on a background field of the $\mathbb Z_{N}^{(1)}$ center symmetry and examine the BC 't Hooft anomaly on $\mathbb R^3\times \mathbb S^1_L$. This can be achieved by recalling the exact same procedure we followed in $4$ dimensions. Here, however, we can entertain the fact that  all  fields are valued in the Cartan subalgebra space, and at energies much smaller than $L^{-1}$ we need to follow the degrees of freedom that enter the semi-classical analysis. We adopt the exact same procedure used in \cite{Poppitz:2020tto} to study the center-symmetry in super Yang-Mills theory. 

To this end,  we enlarge the abelian group $U(1)^{N-1}$  to $U(1)^{N}$ by going to  the $\mathbb R^N$ basis \cite{Argyres:2012ka}. The wights of the defining representation in the $\mathbb R^N$ basis are $\bm \nu_A=\bm e_A-\frac{1}{N}\sum_{A=1}^{N}\bm e_A$, for $A=1,2,...,N$, and $\{\bm e_A\}$ are basis vectors spanning the $\mathbb R^N$ space, while  the simple roots are given by $\bm \alpha_A=\bm e_A-\bm e_{A+1}$, for $A=1,2,...,N$. Let $\tilde F^A$ be the $U(1)^N$ fields in this basis. Then, the periods of  $\tilde F^A$ are given by $\oint \tilde F^A=2\pi \mathbb Z$, where the integration is performed on $2$-cycles. In this basis we have one spurious degree of freedom, which can be eliminated  by imposing the following constraint on the $U(1)^{N}$ fields: $\sum_{A=1}^N \tilde F^A=2\pi n$, for some integer $n$. Everything we have said about $\tilde F^A$ also applies to $\tilde \varphi^A$, the $U(1)^N$ gauge field component along $\mathbb S^1_L$. 

Upon compactifying the theory on $\mathbb S^1_L$, the $4$-dimensional $\mathbb Z_{N}^{(1)}$ symmetry decomposes into a $1$-form symmetry that acts on Wilson's loops on $\mathbb R^3$ (here we need to compactify $\mathbb R^3$ on a large $3$-torus) and a $0$-form symmetry that acts on Polyakov loops wrapping $\mathbb S^1_L$.  The latter vanish in a center-symmetric vacuum $\mbox{tr}_F\left[e^{i \bm \Phi_0\cdot \bm H}\right]=0$, where $\bm H$ are the generators of the Cartan subalgebra. Thus, the background fields of the $\mathbb Z_{N}^{(1)}$ symmetry  decompose as:
\begin{eqnarray}
B^{c(2)}= B_{3D}^{c(2)}+ B_{\mathbb S^1_L}^{c(1)}\wedge \frac{dx^3}{L}\,,\quad
B^{c(1)}= B_{3D}^{c(1)}+ B_{\mathbb S^1_L}^{c(0)} \frac{dx^3}{L}\,,
\label{color background decomp}
\end{eqnarray}
such that the conditions $NB_{3D}^{c(2)}=dB_{3D}^{c(1)}$ and $N  B_{\mathbb S^1_L}^{c(1)}=dB_{\mathbb S^1_L}^{c(0)}$ are obeyed. The various $0$-form and $1$-form fields obey the quantization conditions $\oint_{2-\mbox{cycle}} dB_{3D}^{c(1)}\in 2\pi \mathbb Z$,  $\oint_{1-\mbox{cycles}} dB_{\mathbb S^1_L}^{c(0)}\in \mathbb Z$, $\oint_{2-\mbox{cycle}} B_{3D}^{c(2)}\in \frac{2\pi}{N}\mathbb Z$, $\oint_{1-\mbox{cycles}}  B_{\mathbb S^1_L}^{c(1)}\in \frac{2\pi}{N}\mathbb Z$.

Next, we use  the fact that the $4$-dimensional combination $\tilde F-B^{c(2)}$ is invariant under the $1$-form gauge transformation via the $1$-form field $\lambda^{(1)}$. Thus,  we can write a $3$-dimensional effective field theory, which is invariant under the same $\lambda^{(1)}$ transformation, by  replacing each component of $\bm F$ by $\tilde F^A- B_{3D}^{c(2)}$ and each component of $d\bm \varphi$ by $d\tilde \varphi^A - B_{\mathbb S^1_L}^{c(1)}$ in (\ref{low energy lag}). Thus, we obtain  the bosonic part of the Lagrangian(we suppress $V(\bm \Phi)$):
\begin{eqnarray}
\nonumber
{\cal L}_{3D}^{\scriptsize\mbox{bosonic}}&=&-\frac{L}{4g^2}\sum_{A=1}^{N}\left(\tilde F^A_{\mu\nu}- B_{\mu\nu,3D}^{c(2)}\right)\left(\tilde F^{\mu\nu,A}- B_{3D}^{\mu\nu,c(2)}\right)\\
\nonumber
&& -\frac{b^{(0)}}{8\pi^2}\epsilon^{\alpha\mu\nu}\sum_{A=1}^N\left(\partial_\alpha\tilde \varphi^A  - B_{\alpha,\mathbb S^1_L}^{c(1)}\right) \left(\tilde F_{\mu\nu}^A- B_{\mu\nu,3D}^{c(2)}\right)\\
&&+\frac{1}{2g^2 L}\sum_{A=1}^N \left(\partial_\alpha\tilde \varphi^A  - B_{\alpha,\mathbb S^1_L}^{c(1)}\right) \left(\partial^\alpha\tilde \varphi^A  - B_{\mathbb S^1_L}^{\alpha,c(1)}\right)\,. 
\label{3D bosonic}
\end{eqnarray}
Next, we need to eliminate the spurious degrees of freedom contained in $\tilde F^A$ and $d\tilde \varphi^A$, and in the same time use a duality transformation to write the effective action in terms of the $U(1)^N$ dual photons $\tilde \sigma^A$. Both of these requirements can be implemented using the following auxiliary Lagrangian:
\begin{eqnarray}
\nonumber
{\cal L}^{\scriptsize\mbox{auxilary}}&=&-\frac{1}{4\pi}\sum_{A=1}^N \epsilon^{\lambda \mu\nu}\partial_\alpha \tilde \sigma^A \tilde F^{A}_{\mu\nu}+\frac{1}{4\pi}\epsilon^{\mu\nu \alpha}u_{\alpha}\sum_{A=1}^N \left(\tilde F^A_{\mu\nu}- B_{\mu\nu,3D}^{c(2)}\right)\,,\\
&&+\frac{1}{4\pi}v_{\alpha}\sum_{A=1}^N  \left(\partial^\alpha\tilde \varphi^A  - B_{\mathbb S^1_L}^{\alpha,c(1)}\right)\,,
\end{eqnarray}
where $u_\alpha$ and $v_\alpha$ are the two Lagrange multipliers used to impose the two constraints: 
\begin{eqnarray}
\sum_{A=1}^N \left(\tilde F^A_{\mu\nu}- B_{\mu\nu,3D}^{c(2)}\right)=0\,,\quad \sum_{A=1}^N  \left(\partial^\alpha\tilde \varphi^A  - B_{\mathbb S^1_L}^{\alpha,c(1)}\right)=0\,.
\label{constraints}
\end{eqnarray}
Then, we substitute (\ref{constraints}) into (\ref{3D bosonic}) and vary ${\cal L}_{3D}^{\scriptsize\mbox{bosonic}}+{\cal L}^{\scriptsize\mbox{auxilary}}$ with respect to $\tilde F^A_{\mu\nu}$ to find:
\begin{eqnarray}
\tilde F^A_{\mu\nu}= B_{\mu\nu,3D}^{c(2)}-\frac{g^2}{2\pi L}\epsilon_{\mu\nu \alpha}\left(\partial^\alpha \tilde \sigma^A-u^\alpha +\frac{b^{(0)}}{2\pi}\left(\partial^\alpha \tilde \varphi ^A-\frac{1}{N}\sum_{B=1}^N \partial^\alpha \tilde \varphi ^B\right)\right)\,.
\label{final FA}
\end{eqnarray}
Finally, we substitute (\ref{final FA}) into ${\cal L}_{3D}^{\scriptsize\mbox{bosonic}}+{\cal L}^{\scriptsize\mbox{auxilary}}$ to obtain the dual Lagrangian:
\begin{eqnarray}
\nonumber
{\cal L}^{\scriptsize \mbox{bosonic, dual}}_{3D}&=&\frac{g^2}{8\pi^2 L}\sum_{A=1}^N \left| \partial_\alpha \tilde \sigma^A-\frac{1}{N}\sum_{B=1}^N \partial_\alpha \tilde \sigma ^B+\frac{b^{(0)}}{2\pi}\left(\partial_\alpha \tilde\varphi-\frac{1}{N}\sum_{B=1}^N\partial_\alpha \tilde \varphi^B\right) \right|^2\\
&+&\frac{1}{2g^2 L}\sum_{A=1}^N\left|\partial_\alpha \tilde\varphi^A -B_{\alpha,\mathbb S^1_L}^{c(1)}\right|^2-\frac{1}{4\pi}\sum_{A=1}^N \epsilon^{\alpha \mu\nu}\partial_\alpha\tilde \sigma^A B_{\mu\nu,3D}^{c(2)}\,. 
\label{Lagrangian bosonic dual}
\end{eqnarray}
This is the exact same Lagrangian that was obtained in \cite{Poppitz:2020tto} for super Yang-Mills theory. As we show below, this Lagrangian needs to be augmented with the fermionic part to match the full BC anomaly. 

The last term in (\ref{Lagrangian bosonic dual}) is going to play the main role in what we do next. In terms of differential forms, this term reads: 
\begin{eqnarray}
{\cal L}^{\scriptsize \mbox{bosonic, dual}}_{3D}\supset-\frac{1}{2\pi}\sum_{A=1}^N d\tilde \sigma^A \wedge B_{3D}^{c(2)}\,.
\label{center background term}
\end{eqnarray}
Under a $\mathbb Z^{d\chi}_{2T_{\cal R}}$ transformation  $d\tilde \sigma^A$ and $b^{(0)}$ transform as $d\tilde \sigma^A\rightarrow d\tilde \sigma^A -d\tilde \varphi^A$, $b^{(0)}\rightarrow b^{(0)}+2\pi$ (see (\ref{sigma transformation})), and only the term $-\frac{1}{2\pi}\sum_{A=1}^N d\tilde \sigma^A \wedge B_{3D}^{c(2)}$ contributes to  the variation of ${\cal L}^{\scriptsize \mbox{bosonic, dual}}$:
\begin{eqnarray}
e^{i\delta S^{\scriptsize \mbox{bosonic, dual}}_{3D}}=e^{-\frac{i}{2\pi}\sum_{A=1}^N \int d\tilde \varphi^A \wedge B_{3D}^{c(2)} }\,.
\end{eqnarray}
Then using the second constraint in (\ref{constraints}), $\sum_{A=1}^N d\tilde \varphi^A=NB^{c(1)}_{\mathbb S^1_L}$, along with the quantization conditions of $B_{3D}^{c(2)} $ and $B^{c(1)}_{\mathbb S^1_L}$ we find
\begin{eqnarray}
e^{i\delta S_{3D}^{\scriptsize \mbox{bosonic, dual}}}=e^{-\frac{i2\pi}{N}}\,.
\end{eqnarray}

The above manipulations show that the $\mathbb Z^{(1)}_N$ background lurks deep in the IR and that it couples to the dual photons. This, however, does not capture the full BC anomaly;  we still need to compute the variation of the fermions action in the $\mathbb Z_N^{(1)}$ and $U(1)_B$ backgrounds. This can be obtained from the $U(1)_B$ topological charge density, the second equation in (\ref{topological charge densities}). Substituting (\ref{baryon decomp}) and (\ref{color background decomp}) into (\ref{topological charge densities}) and integrating by parts along the $\mathbb S^1_L$ direction, we obtain the fermion contribution to the variation of the action:
\begin{eqnarray}
\delta S^{\scriptsize \mbox{fermionic}}=\frac{2\pi }{T_{\cal R}}{}\frac{\mbox{dim}_{\cal R}}{4\pi^2} \int \left(dV_{3D}^{(1)}-nB^{(2)}_{3D}\right)\wedge \left(dV^{(0)}_{\mathbb S^1_L}-nB_{\mathbb S^1_L}^{(1)} \right)\,.
\end{eqnarray}
Using the quantization condition $\frac{1}{4\pi^2}\int dV_{3D}^{(1)}\wedge dV_{\mathbb S^1_L}^{(0)}\in \mathbb Z$ along with the quantization conditions of $B^{(2)}_{3D}$ and $B_{\mathbb S^1_L}^{(1)}$ we find:
\begin{eqnarray}
\delta S^{\scriptsize \mbox{fermionic}}=\frac{2\pi \mbox{dim}_{\cal R}}{T_{\cal R}} \left(\frac{n}{N}+\ell\right)^2\,, \ell \in \mathbb Z\,,
\label{fermionic action variation}
\end{eqnarray}
and finally we recover the BC anomaly on $\mathbb R^3\times \mathbb S^1_L$:
\begin{eqnarray}
e^{i\delta S_{3D}^{\scriptsize \mbox{bosonic, dual}}+i\delta S^{\scriptsize \mbox{fermionic}}}=e^{i \frac{2\pi}{T_{\cal R}}\left( T_{\cal R}\left(1-\frac{1}{N}\right)+\mbox{dim}_{\cal R} \left(\frac{n}{N}+\ell\right)^2  \right)}\,,
\label{final expression of BC anomaly on the circle}
\end{eqnarray}
which is exactly (\ref{BC anomaly on R4}), the BC anomaly computed directly on $\mathbb R^4$.

 We conclude the following: 
\begin{itemize}

\item Our analysis shows hat the $\mathbb Z_N^{(1)}$ center acts non-trivially on the dual photons and, when accompanied with the contribution from $U(1)_B$, it produces the correct BC anomaly deep in the IR. This suggests that the BC anomaly is seen and influence the dynamics at all scales.

\item Unlike the BC anomaly, which makes use of the higher-form symmetries, the traditional 't Hooft anomalies are variations of local terms in the action when the theory is  compactified on a small circle. This is clear from the treatment of $\delta S^{\scriptsize \mbox{fermionic}}$ above. Switching off the center background $B^{(2)}_{3D}$ and $B_{\mathbb S^1_L}^{(1)}$, we immediately lose the term (\ref{center background term}) and find $\delta S^{\scriptsize \mbox{bosonic, dual}}=0$  and $ \delta S^{\scriptsize \mbox{fermionic}}=\frac{\mbox{dim}_{\cal R}}{T_{\cal R}}\times \mbox{integer}$. This is exactly the $\mathbb Z_{2T_{\cal R}}^{d\chi}\left[U(1)_B\right]^2$ traditional 't Hooft anomaly. We see right away that this variation of the action is a phase that does not talk to the photons; the dynamics on $\mathbb R^3\times \mathbb S^1_L$ have to obey the BC anomaly, while it is transparent to the traditional $0$-form anomaly. The latter is obeyed by fiat. This observation generalizes the observation that appeared first in  \cite{Poppitz:2020tto}: the cubic- and mixed-$U(1)$ anomalies are matched by local background-field-dependent topological terms instead of chiral-Lagrangian Wess-Zumino-Witten terms and the $1$-form center symmetry talks directly to the dual photons.  

\item It is also important to emphasize, as is well known,  that matching the BC anomaly on $\mathbb R^3\times \mathbb S^1_L$ precludes  a unique gapped vacuum. Such vacuum leaves $\delta \tilde \sigma^A=0$, and hence, $\delta S^{\scriptsize \mbox{bosonic, dual}}=0$, a variations that does not match the anomaly.  Therefore, the anomaly implies that either there exist massless dual photons in the spectrum and/or the discrete chiral symmetry has to break spontaneously, which yields multiple degenerate vacua. We shall see examples of these two possibilities in the following sections. 

\end{itemize}

\subsection{The BC anomaly on $\mathbb R^2\times \mathbb S^1_L\times \mathbb S^1_\beta$}
\label{The BC anomaly at finite T}

In this section we continue our investigation of the BC anomaly as we heat the semi-classical theory that lives on $\mathbb R^3\times \mathbb S^1_L$. Turning on a finite temperature $T$ is equivalent  to compactifying the time direction $x^0$ on a circle $\mathbb S^1_\beta$ of circumference $\beta=\frac{1}{T}$ and giving the fermions anti-periodic boundary conditions on $\mathbb S^1_\beta$. We say that the theory lives on $\mathbb R^2\times \mathbb S^1_L\times \mathbb S^1_\beta$. In order to follow the anomaly from $\mathbb R^3\times \mathbb S^1_L$ to $\mathbb R^2\times \mathbb S^1_L\times \mathbb S^1_\beta$, we decompose the background fields $B^{c(2)}_{3D}$ and $B^{c(1)}_{3D}$ into fields in the $\mathbb R^2$ and $\mathbb S^1_\beta$ directions:
\begin{eqnarray}
B_{3D}^{c(2)}=B_{2D}^{c(2)}+B_{\mathbb S^1_\beta}^{c(1)}\wedge \frac{dx^0}{\beta}\,,\quad B_{3D}^{c(1)}=B_{2D}^{c(1)}+B_{\mathbb S^1_\beta}^{c(0)} \frac{dx^0}{\beta}\,,
\label{reduction on thermal circle}
\end{eqnarray}
such that the constraints $NB^{c(2)}_{2D}=dB_{2D}^{(1)}$ and $NB_{\mathbb S^1_\beta}^{c(1)}=dB_{\mathbb S^1_\beta}^{c(0)} $ are obeyed.  The background fields obey the quantization conditions $\oint dB_{2D}^{(1)}\in  2\pi \mathbb Z$, $\oint B_{2D}^{c(2)}\in\frac{2\pi}{N} \mathbb Z$, $\oint dB_{\mathbb S^1_\beta}^{c(0)}\in 2\pi \mathbb Z$, and $\oint B_{\mathbb S^1_\beta}^{c(1)}\in \frac{2\pi}{N}\mathbb Z$.
At finite temperature we may dimensionally reduce the $3$-dimensional effective field theory down to $2$ dimensions. In particular,  using (\ref{reduction on thermal circle}),  the term (\ref{center background term}), that contains the anomaly, reduces to:
\begin{eqnarray}
{\cal L}_{2D}^{\scriptsize \mbox{bosonic, dual}}\supset-\frac{1}{2\pi}\sum_{A=1}^N d\tilde \sigma^A \wedge B_{\mathbb S^1_\beta}^{c(1)}\,,
\end{eqnarray}
where we have neglected the dual photons derivative in the time direction. Physically, this corresponds to keeping only the zeroth Kaluza-Klein mode of the dual photons and neglecting the higher modes. Under a discrete chiral transformation the dual photons transform as $d\tilde \sigma^A\rightarrow d\tilde \sigma^A -d\tilde \varphi^A$ and the variation of the $2$ dimensional action becomes
\begin{eqnarray}
\delta {\cal L}_{2D}^{\scriptsize \mbox{bosonic, dual}}=\frac{1}{2\pi}\sum_{A=1}^N d\tilde \varphi^A \wedge B_{\mathbb S^1_\beta}^{c(1)}\,.
\end{eqnarray}
Further, we use the second constraint in (\ref{constraints}), $\sum_{A=1}^N d\tilde \varphi^A=N B_{\mathbb S^1_L}^{c(1)}$ , to find the variation of the $2$-dimensional action
\begin{eqnarray}
\delta S_{2D}^{\scriptsize \mbox{bosonic, dual}}=-\frac{N}{2\pi}\int B_{\mathbb S^1_L}^{c(1)}\wedge B_{\mathbb S^1_\beta}^{c(1)}= -\frac{2\pi}{N}\,,
\end{eqnarray}
which is identical to the variation of the $3$-dimensional dual action. This part of the anomaly combines with the contribution from the fermionic action (\ref{fermionic action variation}) to reproduce the BC anomaly (\ref{final expression of BC anomaly on the circle}) at finite temperature. 

The important observation is that the $2$-dimensional dual photons still couple to the $\mathbb Z_N^{(1)}$ center background field, and hence, we expect the anomaly to play a role even at finite temperatures.  Nonetheless, there is an extra layer of complication in $2$ dimensions, thanks to the compact nature of $\bm \sigma$. In $2$ dimensions $\bm\sigma$ have both momentum modes, which are responsible for the logarithmic Coulomb-like force between the monopole instantons, and winding modes. The latter are monodromies of $\bm \sigma$ with a UV cutoff of order $\frac{1}{L}$. These monodromies are the W-bosons and heavy fermions that were not captured by the the low energy effective field theory in $3$ dimensions. As we crank up the temperature and approach the critical temperature of the phase transition/crossover, the heavy excitations  inevitably pop up from vacuum and participate in the dynamics alongside with monopoles and other composite instantons. Eventually, one needs to deal with an electric-magnetic Coulomb gas, which, in general, is a strongly-coupled problem.

In this paper we avoid delving into the anomaly matching in the fully-fledged electric-magnetic Coulomb gas, leaving it for a future investigation. In the next section, however, we give an example that illustrates the idea of the BC anomaly matching at a finite temperature given that we stay well inside the semi-classical weakly-coupled regime. Then, we comment on the fate of this anomaly at very high temperatures.

\section{Examples on $\mathbb R^3\times \mathbb S^1_L$: the $2$-index (anti)symmetric fermions}
\label{Examples on the circle}

In this section we consider several examples on $\mathbb R^3 \times \mathbb S^1_L$ and on  $\mathbb R^2 \times \mathbb S^1_L  \times \mathbb S^1_\beta$  that illustrate the main points of this work: it is the BC anomaly that is responsible for communicating the UV information to  the deep IR. In particular, we found from our analysis in Section \ref{Condensates and role of the BC anomaly} that the BC anomaly is stronger than the traditional $\mathbb Z_{2T_{\cal R}}^{d\chi}\left[U(1)_B\right]^2$ anomaly.  Then, we showed in the previous sections that it is the BC anomaly that couples to the dual photon, and thus, one expects that it controls the breaking pattern of the chiral symmetry.  

\subsection{$SU(4k)$ with  $2$-index symmetric fermions}

\begin{figure}[t] 
   \centering 
	\includegraphics[width=6in]{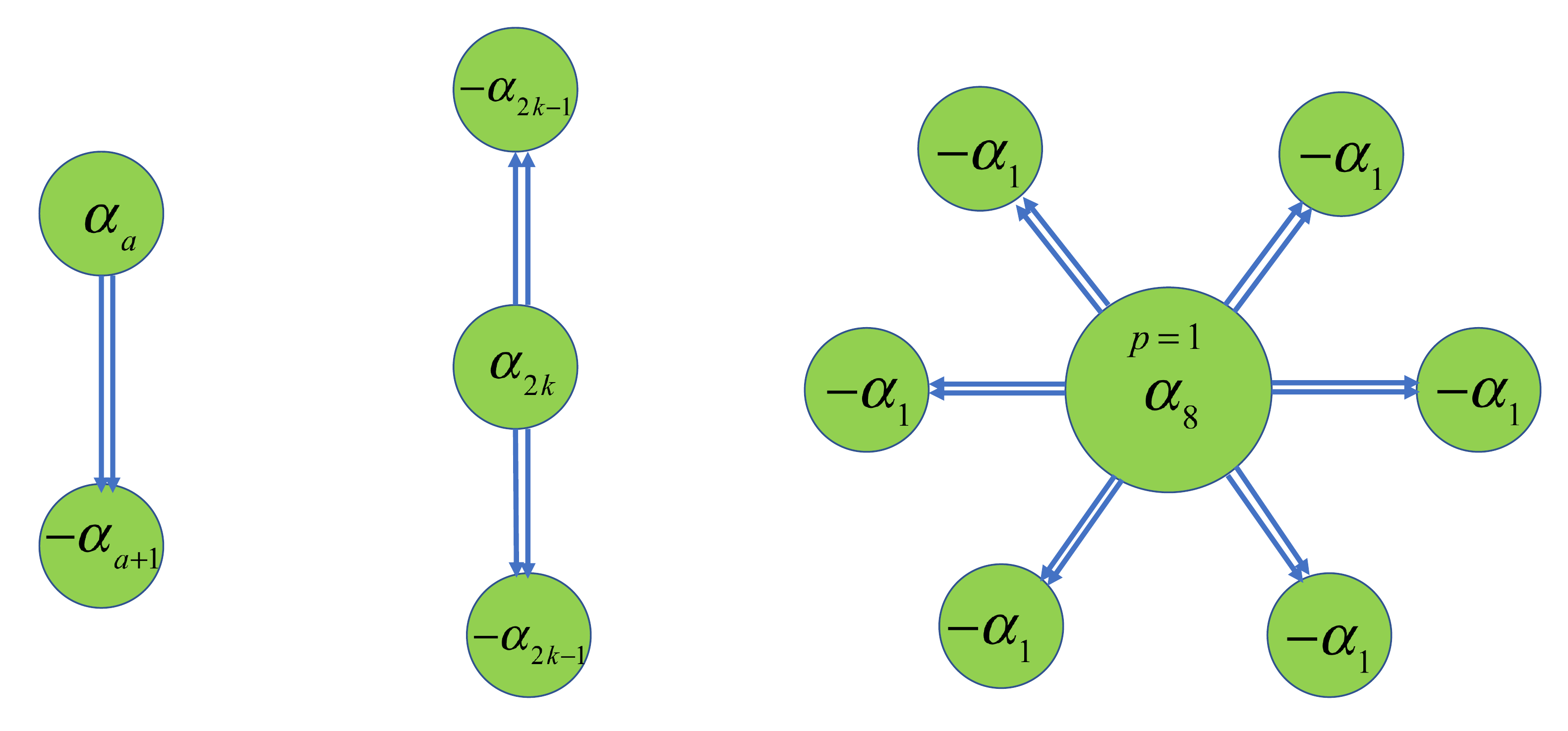}
   \caption{The molecular instantons in the theory on $\mathbb R^3\times \mathbb S^1_L$. From left to right they are the bions, triplets, and higher composites. The latter composite is an example of molecular instanton in $SU(8)$ with fermions in the $2$-index antisymmetric representation. It consist of a $p=1$ Kaluza-Klein monopole attached to $6$ lowest-order monopoles. The $12$ zero modes of the central monopole are soaked up by the orbital monopoles (moons). There is a repulsive Coulomb force between the central monopole and the its moons, which is balanced by the attractive force due to the exchange of the zero modes. The moons, on the other hand, repel each other since they are all charged under the same root, and thus, they are stabilized under this repulsive force.}
   \label{instanton fig}
\end{figure}

We work in the center-symmetric vacuum $\bm \Phi_0=\frac{2\pi \bm \rho}{4k}$, which can be attained by using a double-trace deformation.  The monopole vortices are given by (see (\ref{Monopole vortices}); here we neglect the holonomy fluctuations $\bm \varphi$ and set $b^{(0)}=0$): 
\begin{eqnarray}
\nonumber
&&{\cal M}_a=e^{-\frac{2\pi^2}{k g^2}}e^{i \bm \alpha_a\cdot\bm \sigma}\psi\tilde \psi\,, a\neq 2k\, \mbox{or}\, 4k\,, \quad {\cal M}_{2k}=e^{-\frac{2\pi^2}{k g^2}}e^{i \bm \alpha_{2k}\cdot\bm \sigma}\left(\psi\tilde \psi\right)^2\,,\\
 &&\quad  {\cal M}_{4k}=e^{-\frac{2\pi^2}{k g^2}}e^{i \bm \alpha_{4k}\cdot\bm \sigma}\left(\psi\tilde \psi\right)^2\,.
\end{eqnarray}
Since all the monopoles are dressed with fermion zero modes, they cannot lead to confinement or breaking of the chiral symmetry. Yet, molecular instantons that are composed of two monopoles (bions) \cite{Unsal:2007jx,Anber:2011de} or three monopoles (triplets) \cite{Poppitz:2009uq} can form, see Figure \ref{instanton fig}.  The stability of these molecules is attributed to the fact that the total potential seen by the two or three monopoles admits a stable equilibrium point. This is ascribed to the competition between the repulsive Coulomb force from the dual photons and the attractive force from the exchange of the fermion zero modes (we say that the fermions zero modes are soaked up). In particular, notice that $\bm \alpha_a\cdot \bm\alpha_b=2\delta_{a,b}-\delta_{a,b+1}-\delta_{a,b-1}$, and therefore, only monopole and anti-monopole  that are charged under  neighboring  simple roots can feel the repulsive Coulomb force.  The  bions and triplets with the lowest fugacities are:
\begin{eqnarray}
\nonumber
&&{\cal M}_a\overline{{\cal M}_{a+1}}\,, 1\leq a<2k-1\,\mbox{or}\, 2k+1\leq a<4k-1\, \,,  \quad {\cal M}_{2k}\left(\overline{{\cal M}_{2k-1}}\right)^2\,, \quad  {\cal M}_{2k}\left(\overline{{\cal M}_{2k+1}}\right)^2\,,\\
&& {\cal M}_{2k}\left(\overline{{\cal M}_{2k-1}}\right)\left(\overline{{\cal M}_{2k+1}}\right)\,,\quad
 {\cal M}_{4k}\left(\overline{{\cal M}_{4k-1}}\right)^2\,, \quad  {\cal M}_{4k}\left(\overline{{\cal M}_{1}}\right)^2\,,\quad {\cal M}_{4k}\left(\overline{{\cal M}_{1}}\right)\left(\overline{{\cal M}_{4k-1}}\right)\,,
\end{eqnarray}
as well as their anti-bions and anti-triplets. The proliferation of bions and triplets generates a potential of $\bm \sigma$:
\begin{eqnarray}
\nonumber
V(\bm\sigma)&=&-e^{-\frac{6\pi^2}{kg^2}}\left\{\cos\left(\bm \alpha_{2k}-2\bm \alpha_{2k-1}\right)\cdot \bm \sigma+\cos\left(\bm \alpha_{2k}-2\bm \alpha_{2k+1}\right)\cdot \bm \sigma
+\cos\left(\bm \alpha_{4k}-2\bm \alpha_{4k-1}\right)\cdot \bm \sigma\right.\\
\nonumber
&&\left.+\cos\left(\bm \alpha_{4k}-2\bm \alpha_1\right)\cdot \bm \sigma+\cos\left(\bm \alpha_{2k}-\bm \alpha_{2k-1}-\bm \alpha_{2k+1}\right)\cdot \bm \sigma+\cos\left(\bm \alpha_{4k}-\bm \alpha_{4k-1}-\bm \alpha_1\right)\cdot \bm \sigma\right\}\\
&&-e^{-\frac{4\pi^2}{kg^2}}\sum_{\{1\leq i<2k-1\}\cup\{2k+1\leq i<4k-1\}}\cos\left(\bm \alpha_{i}-\bm \alpha_{i+1}\right)\cdot \bm \sigma\,.
\label{effective potential 2-index symmetric}
\end{eqnarray}
The triplets fugacity is exponentially suppressed compared to the bions fugacity and one might be tempted to ignore the triplets. This, however, leaves some flat directions, i.e., massless photons\footnote{Note, however, that massless photons can still match the BC anomaly.}, which are lifted once we take the triplets into account.   

One can easily check that the potential admits a global minimum at $\bm \sigma=0$, and then we can use the chiral transformation $\bm \sigma \rightarrow \bm \sigma-\bm C$, where $\bm C$ is given by (\ref{C}), to obtain the rest of the vacua:
\begin{eqnarray}
\bm \sigma_0=\frac{2\pi n}{4k+2}\left( 2 \bm w_{2k}+\sum_{a=1, a\neq 2k}^{4k-1}\bm w_a\right)\,,\quad n=0,1,..., 4k+1.
\end{eqnarray}
As promised, there are $4k+2$ distinct vacua, which are required to match the BC anomaly.

\subsection{$SU(4k)$ with  $2$-index antisymmetric fermions}

We also work in the center-symmetric vacuum. The monopole vertices are given by:
\begin{eqnarray}
\nonumber
&&{\cal M}_a=e^{-\frac{2\pi^2}{k g^2}}e^{i \bm \alpha_a\cdot\bm \sigma}\psi\tilde \psi\,, a\neq 2k\, \mbox{or}\, 4k\,, \quad {\cal M}_{2k}=e^{-\frac{2\pi^2}{k g^2}}e^{i \bm \alpha_{2k}\cdot\bm \sigma}\,,\quad
 {\cal M}_{4k}=e^{-\frac{2\pi^2}{k g^2}}e^{i \bm \alpha_{4k}\cdot\bm \sigma}\,,
\end{eqnarray}
while the bions are
\begin{eqnarray}
{\cal M}_a\overline{{\cal M}_{a+1}}\,,\quad   1\leq a<2k-1\,\mbox{or}\, 2k+1\leq a<4k-1\,.
\end{eqnarray}
The proliferation of the bions and the two monopoles ${\cal M}_{2k}$ and ${\cal M}_{4k}$ leaves flat directions, and in order to lift them one needs to take into account higher Kaluza-Klein monopoles. 
 
Before discussing these higher order corrections, one  wonders about the possibility of the formation of  bion-like compositions between not neighboring monopoles, e.g., bions of the form ${\cal M}_{a}\overline{\cal{M}}_{a+2}$ that could lift the flat directions.  The problem, though, is that such compositions are unstable against the attractive force due to the exchange of fermions zero modes. Also, the absence of any kind of Coulomb interactions between the monopoles (remember that $\bm \alpha_{a}\cdot \bm \alpha_{a+2}=0$) eleminates the possibility of analytically continuing the coupling constant $g$, i.e., sending $g\rightarrow -g$, that could generate a repulsive coulomb force to compete with the fermion attractive force. This is the famous Bogomolny Zin-Justin analytical continuation prescription that has been used in several works  to stabilize bion-like objects, see, e.g., \cite{Poppitz:2012sw}.  In summary, we do not expect bions of the type ${\cal M}_{a}\overline{\cal{M}}_{a+2}$ to form in vacuum. 

Now, we need to go to the next-to-next-to-leading order in fugacity and consider the higher Kaluza-Klein monopoles (\ref{composite monopole}). A typical example of a complex molecule that can lift the flat directions is composed of a $p=1$ Kaluza-Klein monopole charged under $\bm\alpha_{4k}$, which has a total of $8k-4$ fermion zero modes, and $4k-2$ anti-monopoles charged under the root $-\bm\alpha_{1}$:   
\begin{eqnarray}
{\cal M}_{4k}^{p=1}\left[\overline{{\cal M}_{1}} \right]^{4k-2}\,,
\end{eqnarray}
see Figure \ref{instanton fig}.
The proliferation of the monopoles, bions, and higher composites generates masses for all photons and leads to the full breaking $\mathbb Z_{8k-4}^{d\chi}\rightarrow \mathbb Z_2$. The theory admits $4k-2$ distinct vacua:
\begin{eqnarray}
\bm \sigma_0=\frac{2\pi n}{4k-2}\sum_{a=1, a\neq 2k}^{4k-1}\bm w_a\,,\quad n=0,1,..., 4k-3.
\end{eqnarray}
%

\subsection{The BC anomaly at finite temperature}

In this section we attempt to partially answer  the question about the BC anomaly matching at finite temperature. As we pointed out in Section \ref{The BC anomaly at finite T}, we can reduce the problem to $2$ dimensions by compactifying the time direction on a circle and keeping only the zero mode of the dual photons. Definitely, if the temperature is high enough, then the W-bosons and heavy fermions will be liberated and their effects, in addition to the monopoles and composite instantons, have to be taken care of. The problem, then, reduces to an electric-magnetic Coulomb gas, which in general is a strongly-coupled system. This Coulomb gas was considered before in the $SU(2)$ and $SU(3)$ cases with adjoint fermions, see \cite{Anber:2011gn,Anber:2012ig,Anber:2013xfa,Anber:2013doa,Anber:2018ohz}. Non of these works, however, addressed the issue of anomaly matching. Here, we do not provide a full solution to the anomaly-matching problem at all temperatures, which will be pursued somewhere else. Let us, at least, show how  the BC anomaly is being matched as we crank up the temperature and stay in the weakly-coupled regime. We comment on the fate of the BC anomaly at very high temperatures at the end of the section.

From here on we work in $2$ dimensions. The general structure of $V(\bm \sigma)$   takes the form of a collection of cosine terms, see e.g., (\ref{effective potential 2-index symmetric}), $V(\bm \sigma)=\sum_{m}y_m\cos (\bm Q_m \cdot\bm \sigma)$, where $y_m$ is the fugacity of the instanton, $\bm Q_m$, is its charge, and the sum runs over the various instanton types: monopoles, bions, etc. One, then,  expands the cosine terms and write the grand canonical partition function as:
\begin{eqnarray}
{\cal Z}=\sum_{m}\sum_{N_m=0}^{\infty} \frac{y_m^{N_m}}{N_m!}\int \prod_{j_m=1}^{N_m} d^2 \vec x_{j_m} e^{-\int {\cal L}_2}\,,
\end{eqnarray}
where
\begin{eqnarray}
{\cal L}_2=\frac{g^2}{8\pi^2 LT}\partial_i \bm \sigma\cdot\partial_i \bm \sigma+\sum_{m}\bm J_m \cdot \bm \sigma\,.
\label{sigma Lag 2D}
\end{eqnarray}
The latin letter $i=1,2$ labels the $\mathbb R^2$ space and  $\bm J_m=\bm Q_m\delta^{(2)}(\vec x-\vec x_m)$ is the current source of an instanton of charge $\bm Q_m$ located at $\vec x_m$.  Then, we can solve the Gaussian system, ignoring the monodromies of $\bm\sigma$ since they correspond to heavy electric excitations not accessible at low temperature, to find the potential energy between two sources:
\begin{eqnarray}
V(\vec x_1,\vec x_2)=-\frac{4\pi LT}{g^2}\bm Q_1\cdot \bm Q_2 \log T|\vec x_1-\vec x_2|\,.
\end{eqnarray}
Next, we substitute  this result into (\ref{sigma Lag 2D}) to obtain the grand canonical partition function of a magnetic Coulomb gas:
\begin{eqnarray}
{\cal Z}=\sum_{m}\sum_{N_m=0}^{\infty} \frac{y_m^{N_m}}{N_m!}\int \prod_{j_m=1}^{N_m} d^2 \vec x_{j_m} e^{ \frac{4\pi LT}{g^2}\sum_{m,m'}\sum_{a\neq b}\bm Q_m^a\cdot \bm Q_{m'}^b \log T|\vec x_a-\vec x_b|}\,,
\end{eqnarray}
and we need to impose a neutrality condition on the gas to avoid IR divergences.  In order to understand what happens as we increase the temperature, we need to follow the fugacities of the magnetic charges under the renormalization group flow. Let us consider a pair of magnetic charges $\bm Q_m$ and $-\bm Q_m$ located at $\vec x_1$ and $\vec x_2$ and separated by a distance $L$. The  pair's contribution to the partition function is
\begin{eqnarray}
\left(\frac{y_m(a)}{a^2}\right)^2\int d^2 \vec x_{1} d^2 \vec x_{2}\left|\frac{\vec x_1-\vec x_2}{a}\right|^{-\frac{4\pi LT}{g^2} \bm Q_m \cdot \bm Q_m}=y_m^2(a)\left(\frac{L}{a}\right)^{4{-\frac{4\pi LT}{g^2} \bm Q_m \cdot \bm Q_m}}\,,
\end{eqnarray}
where $a$ is a UV cutoff. Demanding the invariance of the left hand side under the renormalization group flow means:
\begin{eqnarray}
y_m^2(a)\left(\frac{L}{a}\right)^{4{-\frac{4\pi LT}{g^2} \bm Q_m \cdot \bm Q_m}}=y_m^2(ae^b)\left(\frac{L}{ae^b}\right)^{4{-\frac{4\pi LT}{g^2} \bm Q_m \cdot \bm Q_m}}\,.
\end{eqnarray}
Taking the derivative with respect to $b$ and setting $b=0$, we obtain the renormalization group equations of the fugacities
\begin{eqnarray}
\frac{dy_m}{db}=\left(2-\frac{2\pi LT}{g^2}\bm Q_m \cdot \bm Q_m\right)y_m\,.
\label{RG equation magnetic}
\end{eqnarray}
%

\begin{figure}[t] 
   \centering 
	\includegraphics[width=6in]{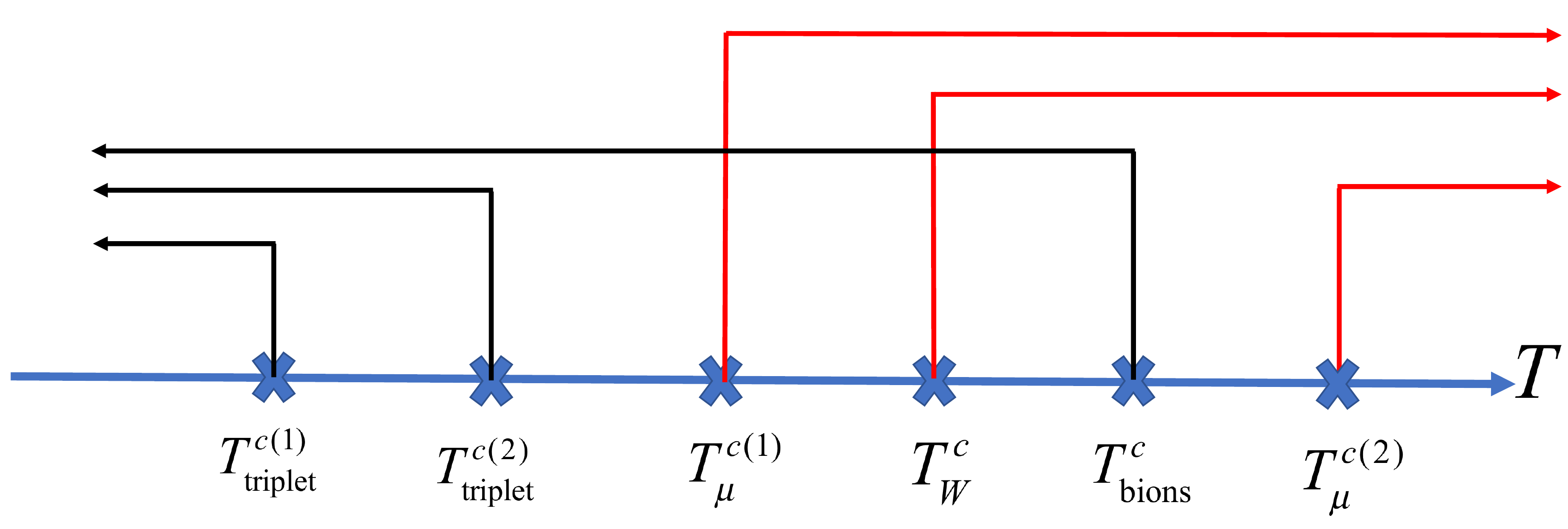}
   \caption{The critical temperatures of $SU(8)$ with fermions in the $2$-index symmetric representation. The directions of the arrows indicate the temperature range at which the fugacity of the corresponding charge becomes relevant. The domination of magnetic (electric) charges is indicated by black (red) arrows.  See the text for a detailed discussion.}
   \label{transition}
\end{figure}

Equation (\ref{RG equation magnetic}) determines the critical temperature above which the fugacity of a certain magnetic charge becomes irrelevant: 
\begin{eqnarray}
T^c_m=\frac{g^2}{\pi L \bm Q_m\cdot \bm Q_m}\,.
\end{eqnarray}
Therefore, as we heat the system, magnetic charges with bigger $\bm Q_m\cdot \bm Q_m$ decouple first. This is the Berezinskii-Kosterlitz-Thouless (BKT) transition. 

In oder to make sure that $T^c_m$ is well within the semi-classical regime---so that we can neglect the effect of the electric charges, hence, the renormalization group analysis we performed above is justified---we need to compute the critical temperatures at which the electric excitations, the W-bosons and heavy fermions, dominate the plasma. An electric charge with mass $M$ will have a fugacity given by the Boltzmann factor $y_e=e^{-\frac{M}{T}}$, and the electric potential between two charges is given by:
\begin{eqnarray}
V(\vec x_1,\vec x_2)=-\frac{g^2}{4\pi LT}\bm Q_1\cdot \bm Q_2 \log T\left|\vec x_1-\vec x_2\right|\,.
\end{eqnarray}
Then, we can repeat the above steps to find the renormalization group equations of the electric fugacities: 
\begin{eqnarray}
\frac{dy_e}{db}=\left(2-\frac{g^2}{8\pi LT}\bm Q_e \cdot \bm Q_e\right)y_e\,,
\label{RG equation electric}
\end{eqnarray}
from which we find the critical temperature above which the electric charges proliferate:
\begin{eqnarray}
T^c_e=\frac{g^2}{16\pi L}\bm Q_e \cdot \bm Q_e\,.
\end{eqnarray}
As expected, the bigger the electric charge $\bm Q_e \cdot \bm Q_e$, the higher the critical temperature above which it dominates the plasma, which is the exact opposite of the magnetic critical temperature. Staying inside the semi-classical, magnetically disordered, regime demands $T_c^m<T_c^e$.

As an example, let us apply this treatment to $SU(8)$ with fermions in the $2$-index symmetric representation. This theory contains two types of magnetic charge: the bions, that carry  charge $Q=\bm \alpha_a-\bm \alpha_{a+1}$, $a=1,2,5,6$, and triplets. There are also two types of triplets:  the first type, e.g., $ {\cal M}_{4}\left(\overline{{\cal M}_{3}}\right)^2$ has charge $\bm Q=\bm \alpha_4-2\bm \alpha_3$, and the second type, e.g., ${\cal M}_{4}\left(\overline{{\cal M}_{3}}\right)\left(\overline{{\cal M}_{5}}\right)$   has charge $\bm Q=\bm \alpha_4-\bm \alpha_3-\bm\alpha_5$. Using the renormalization group equation of the magnetic fugacities (\ref{RG equation magnetic}), we find $3$ distinct critical temperatures:
\begin{eqnarray}
T^{c(1)}_{\scriptsize\mbox{triplet}}=\frac{g^2}{14\pi L}\,, \quad T^{c(2)}_{\scriptsize\mbox{triplet}}=\frac{g^2}{11\pi L}\,, \quad T^{c}_{\scriptsize\mbox{bion}}=\frac{g^2}{6\pi L}\,,
\end{eqnarray}
which correspond, respectively, to the temperatures above which the first triplet, the second triplet, and then the bions become irrelevant. Similarly, we use the weights of the $2$-index symmetric representation, the fact that the W-bosons carry charges valued in the root lattice,  along with the renormalization group equations of the electric fugacities to find $3$ distinct critical temperatures: 
\begin{eqnarray}
T^{c(1)}_{\bm \mu}=\frac{3g^2}{32\pi L}\,,\quad T^{c}_{W}=\frac{g^2}{8\pi L}\,,\quad T^{c(2)}_{\bm \mu}=\frac{7g^2}{32\pi L}\,,
\end{eqnarray}
which correspond, respectively, to the temperatures at which a first group of heavy fermions, the W-bosons, and then a second group of heavy fermions become relevant. 

The $6$ critical temperatures and the corresponding relevant excitations are depicted in Figure \ref{transition}. At temperatures smaller than $T^{c(1)}_{\scriptsize\mbox{triplet}}$ the chiral symmetry is fully broken and all the photons are massive.  For temperatures in the range $T^{c(1)}_{\scriptsize\mbox{triplet}}<T<T^{c(2)}_{\scriptsize\mbox{triplet}}$ the first type of triplets decouple leaving behind a vacuum with one flat direction, i.e., a single massless photon. This can be envisaged by studying the effective potential (\ref{effective potential 2-index symmetric}) after neglecting the first type of triplets.  Then, as we crank up the temperature to the range $T^{c(2)}_{\scriptsize\mbox{triplet}}<T<T^{c(1)}_{\bm \mu}$ the second type of triplets decouple leaving behind $3$ massless photons. Interestingly, as long as the temperature is below $T^{c(1)}_{\bm \mu}$, the theory is still inside the semi-classical, magnetically disordered, domain and the BC anomaly  is always matched either by the multiple vacua or by the massless photons. In this range of temperatures  the BC anomaly is not local in the sense that it is felt at arbitrarily long distances. 

For temperatures above $T^{c(1)}_{\bm \mu}$ the electrically confined charges are liberated and it becomes harder to analyze the system, a study that is left for the future. We recall that the theory at hand has a genuine $\mathbb Z^{(1)}_2$ $1$-form symmetry acting on the Polyakov's loops on $\mathbb R^3$.  We expect a confinement/deconfinement phase transition to occur in the temperature range $T^{c(1)}_{\bm \mu}<T<T^c_{\scriptsize\mbox{bion}}$. Presumably this is a first order transition given the large number of degrees of freedom\footnote{It was found in multiple studies that the transition is first order for $SU(N\geq 3)$, \cite{Anber:2012ig,Anber:2014lba}.}. Beyond the transition temperature the magnetic charges become confined (irrelevant). Since it is the dual photons that lead to the long-range force between monopoles, the fact that the magnetic charges become confined above the phase transition temperature means that the BC anomaly becomes local; it is an overall phase in the transition function that is now matched by fiat, but otherwise does not dictates the dynamics in the deep IR. We also expect the discrete chiral symmetry to be restored above the phase transition temperature.

{\flushleft \bf Acknowledgments:} I would like to thank Erich Poppitz for comments on the manuscript. This work is supported by NSF grant PHY-2013827. 
\bibliography{References}

\bibliographystyle{JHEP}

\end{document}